\documentclass[preprint,review]{elsarticle}

\journal{Journal of \LaTeX\ Templates}

\usepackage[ansinew]{inputenc}
\usepackage{natbib}
\usepackage[section]{placeins}
\usepackage{color,placeins}
\usepackage{graphicx,amsmath}
\usepackage{graphics} % for pdf, bitmapped graphics files
\usepackage{epsfig,epstopdf} % for postscript graphics files
\usepackage{mathptmx} % assumes new font selection scheme installed
\usepackage{times} % assumes new font selection scheme installed
\usepackage{amssymb}  % assumes amsmath package installed
\usepackage{amsfonts}
\usepackage{psfrag}
\usepackage{graphicx}
\usepackage{float}
\usepackage{epstopdf}
\usepackage{setspace}
\usepackage{adjustbox}
\usepackage[colorlinks=true,linkcolor=blue]{hyperref}
\usepackage{url}
\usepackage[font=small,skip=5pt]{caption}
\usepackage{subcaption}
\DeclareGraphicsExtensions{.pdf,.png,.jpg}
\usepackage{microtype}
\usepackage{adjustbox}
\usepackage{color}
\usepackage{xcolor}
\usepackage{multirow}
\usepackage{graphicx}
\usepackage{caption}
\usepackage{subcaption}
\usepackage{amssymb,mathtools}
\usepackage{xfrac}
\usepackage{enumerate}
\usepackage{comment}
\hypersetup{
    colorlinks = true,
    linkcolor={red},
    citecolor={magenta},
    urlcolor={green},
    allcolors={blue}
}
\newtheorem{theorem}{Theorem}
\newtheorem{proof}{Proof}

\newtheorem{remark}{Remark}
\newtheorem{assumption}{Assumption}
\newtheorem{case}{Case}
\newtheorem{prop}{Property}

\begin{document}

\begin{frontmatter}

\title{State Estimation-Based Robust Optimal Control of Influenza Epidemics in an Interactive Human Society}

\author{Vahid Azimi}
\address{Department of Energy Resources Engineering, Stanford
	University, Stanford, CA, USA}
%\email{vazimi@stanford.edu}

\author{Mojtaba Sharifi}
\address{Department of Electrical and Computer Engineering, University of Alberta, Edmonton, Canada}
%\email{sharifi3@ualberta.ca}

\author{Seyed Fakoorian}
\address{Department of Electrical Engineering and Computer Science, Cleveland State University,	Cleveland, OH, USA}
%\email{s.fakoorian@csuohio.edu}

\author[TDT]{Thang Nguyen\corref{mycorrespondingauthor}}
\cortext[mycorrespondingauthor]{Corresponding author}
\ead{nguyentienthang@tdtu.edu.vn}

\author[TDT]{Van Van Huynh}
\address[TDT]{Modeling Evolutionary Algorithms Simulation and Artificial Intelligence, Faculty of Electrical \& Electronics Engineering, Ton Duc Thang University, Ho Chi Minh City, Vietnam}

\begin{abstract}
This paper presents a state estimation-based robust optimal control strategy for influenza epidemics in an interactive human society in the presence of modeling uncertainties. Interactive society is influenced by random entrance of individuals from other human societies whose effects can be modeled as a non-Gaussian noise. Since only the number of exposed and infected humans can be measured, states of the influenza epidemics are first estimated by an extended maximum correntropy Kalman filter (EMCKF) to provide a robust state estimation  in the presence of the non-Gaussian noise. An online quadratic program (QP) optimization is then synthesized subject to a robust control Lyapunov function (RCLF) to minimize susceptible and infected humans, while minimizing and bounding the rates of vaccination and antiviral treatment. The joint QP-RCLF-EMCKF meets multiple design specifications such as state estimation, tracking, pointwise control optimality, and robustness to parameter uncertainty and state estimation errors that have not been achieved simultaneously in previous studies. The uniform ultimate boundedness (UUB)/convergence of error trajectories is guaranteed using a Lyapunov stability argument. Simulation results show that the proposed approach achieves appropriate tracking and state estimation performance with good robustness on the influenza epidemics of an interactive human society with population of 16000.
\end{abstract}


\end{frontmatter}

%\linenumbers

%%%%%%%%%%%%%%%%%%%%%%%%%%%%%%%%%%%%%%%%%%%%%%%%%%%%%%%%%%%%%%%%%%%%%%%%%%%%%%%%
\section{Introduction} \label{section.introduction}

Influenza viruses can cause epidemic human diseases that are currently a worldwide health concern. Proper control of influenza epidemics is a crucial task that can mitigate economic and epidemiological burdens. Recent years have witnessed numerous studies in analysis, modeling, and control of influenza epidemiological systems~\cite{Mode1Flu,Sharifi,Opt1Flu,Opt2Flu,FlightFlu,Inf-SMC}. 
Mathematical model of influenza epidemics can provide an opportunity to design model-based control strategies and to analyze the stability of closed-loop systems. Several mathematical models have been proposed for influenza epidemic systems~\cite{Mode1Flu,Mode2Flu,Sharifi}.
In~\cite{Mode1Flu}, compartmental models of the influenza were proposed
while considering the vaccination and antiviral treatment as control inputs. In~\cite{Mode2Flu},
influenza dynamics were modeled by a set of nonlinear differential equations.
In~\cite{Sharifi}, a nonlinear SEIAR model
of the influenza with two control inputs and five states was described.
In this model, the positive state variables S, E, I, A, and R are the Susceptible,
Exposed, Infected, Asymptomatic, and Recovered individuals while rates of vaccination and antiviral treatment are considered positive control inputs.
\begin{figure}[t]
\centering
\vspace{-0em}
\begin{subfigure}[b]{0.17\textwidth}
\hspace*{-7em}
\includegraphics[scale=0.07]{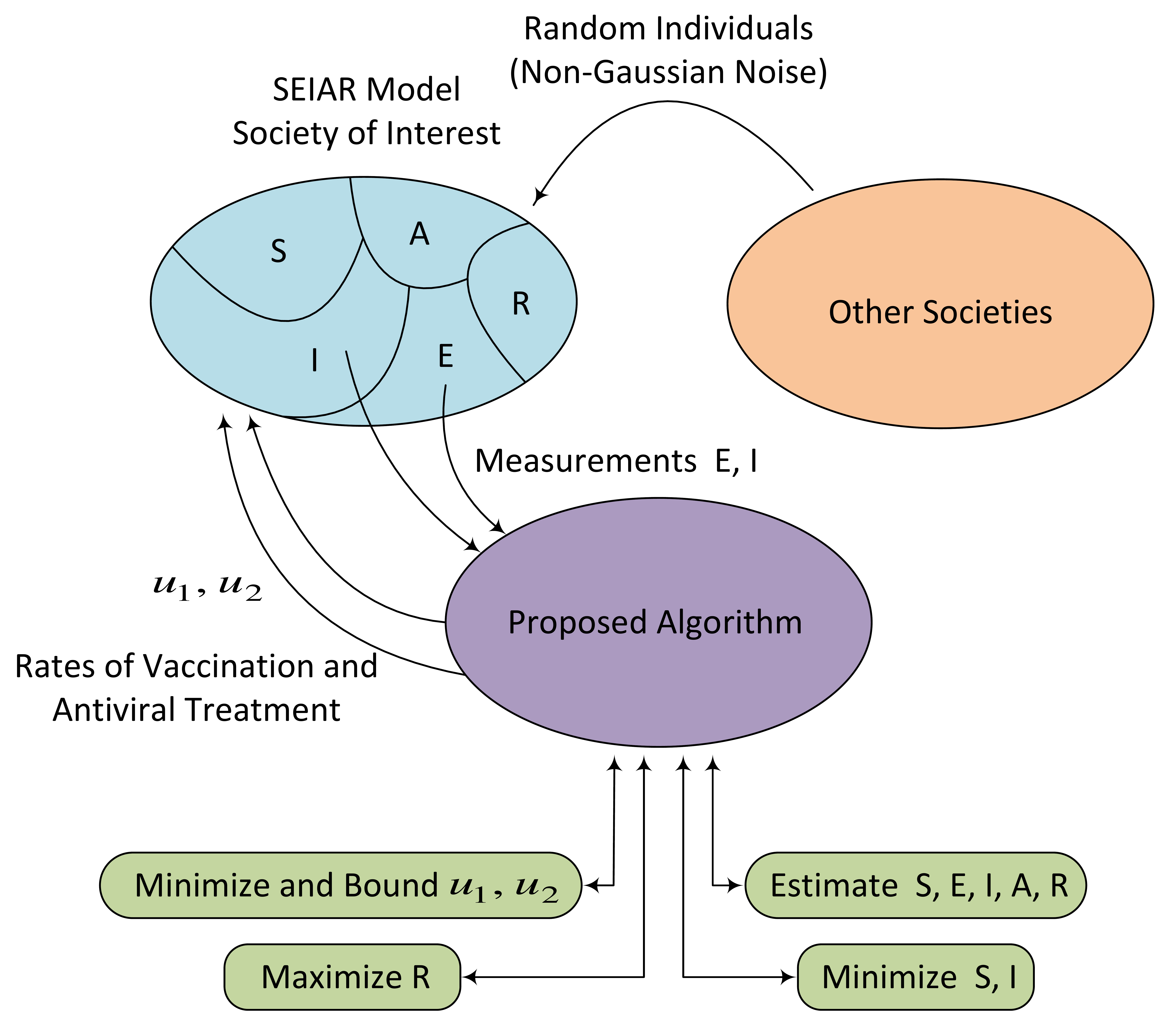}
\end{subfigure}%
\caption{Proposed structure for estimation and control of influenza epidemics
in an interactive human society. Influenza dynamics
are modelled by an SEIAR model with Susceptible (S), Exposed (E),
Infected (I), Asymptomatic (A), and Recovered (R) individuals.
 An interactive society is influenced
by other human societies whose effects on the interactive society
can be modeled as a non-Gaussian noise.} \label{fig.FluFlo}
\vspace{-1em}
\end{figure}

To recover all individuals of a society,
the best intervention strategy is desired to be designed for the influenza epidemics.
Optimal control is one of the widely-used approach that has been
employed to determine
the treatment strategies~\cite{Opt1Flu,Opt2Flu,Opt3Flu,Opt4Flu,Opt5Flu}.
In~\cite{Opt1Flu}, an optimal control problem was employed to minimize the number of infected
individuals at minimal efforts of the vaccination. Different optimal control strategies were suggested in~\cite{Opt2Flu} to minimize the impact of influenza pandemics involving antiviral
treatment and/or the isolation measures. In~\cite{Opt3Flu}, prevention of the pandemic influenza was
enhanced towards evaluating time-dependent optimal prevention policies and considering its
execution cost. In~\cite{Opt4Flu}, a dynamic model of an influenza pandemic model was formulated with the existence of vaccination and treatment, and then analyzed in terms of the vaccine intake variations.
In~\cite{Opt5Flu}, a prioritization scheme for allocation of a sizeable quantity
of influenza vaccine and antiviral drug was described for a stratified population.

\begin{table*}[t]
\centering
\begin{tabular}{|cc|}
\hline
\textbf{Abbreviations} & \\
& \\
MMSE  & Minimum mean square error\\
EKF  &  Extended Kalman filter\\
UKF  & Unscented Kalman filter \\
SEIAR  & Susceptible, exposed, infected, asymptomatic, and recovered\\
PWMC  & Pointwise min-norm control\\
EMCKF & Extended maximum correntropy Kalman filter\\
QP  &  Quadratic programming \\
ES-CLF  & Exponentially stabilizing control Lyanpunov function\\
RCLF  &  Robust control Lyapunov function\\
UUB   & Uniform ultimate boundedness \\
\hline
\end{tabular}
\end{table*}

Note that the above-mentioned optimal control strategies were formulated
with the assumption of fully-known dynamic terms and parameters. However, mathematical models
of the influenza epidemics may contain modeling uncertainties that should be taken into account
in the control design structure. In~\cite{ParaFlu}, a least squares method was employed to estimate unknown parameters of two influenza epidemic models. Although the
estimation performance was validated, no any control strategy was designed to minimize the
infected population. In~\cite{Sharifi}, a robust adaptive sliding mode
 controller was designed for a nonlinear SEIAR model of the influenza
in the presence of parametric uncertainties. In that work,
convergence of susceptible and infected humans to zero was provided by tracking some descending scenarios.
Two robust terms were also incorporated in their devised controller whose gains were updated
using adaptation laws to compensate for the parameter uncertainties.
Stability of closed-loop influenza epidemic system was then proved using
a Lyapunov framework and the Barbalat's lemma.

However, that recent paper~\cite{Sharifi} suffers from several drawbacks.
(i) The main one is that the controller requires accurate measurement of state variables,
while only the population of exposed and infected humans can be measured in practice. (ii) In that approach,
studied human society was assumed to be isolated from other societies.
However, a random entrance of individuals from other
societies into the main society of interest
results in degrading the control performance. This kind of society is called
"\textit{interactive society}" and the effects from the other societies
can be modelled as a non-Gaussian noise as shown in Fig.~\ref{fig.FluFlo}.
   (iii) In their method, although the convergence of system solutions was obtained
and the robustness of closed-loop systems against parametric uncertainties was demonstrated, control optimality, as an important design specification, has not been taken into account. In other words,
tracking, robustness, and
minimizing the rates of vaccination and antiviral treatment should be achieved
at the same time by devising an appropriate control strategy.
(iv) In the normalized SEIAR model, the control signals should be always positive and less than 1.
However, the approach in~\cite{Sharifi} was not able to bound the rates of vaccination and antiviral treatment
in the controller implementation while facing with high parameter
uncertainties
and disturbances. It should also be noted that the rest of the above-mentioned papers suffer
from the shortcomings mentioned in Items (i) and (iv).

The Kalman filter~\cite{KAL} is still the most common
method for state estimation of linear systems because
of its optimality and simplicity. However since the
mathematical model of the influenza comprises a set of nonlinear differential equations, the extension
of Kalman filters, namely the extended Kalman filter (EKF), the unscented Kalman filter (UKF), and particle filters~\cite{Reif2000,sarkka2007unscented,IS3,IET1,IS1,wash1,vahidJPC,UKF-NEW,IS2},
can be alternatively used for the state estimation purpose.
A Kalman filter is derived based on the minimum mean square error (MMSE) criterion, which follows that it uses only second-order information of the signal and it is optimal until the gaussianity of noises is preserved.
However, in this paper, the performance of the ordinary EKF may break down for the influenza epidemics of an interactive society that is disturbed by non-Gaussian noise (when the society is not isolated).
To solve this issue, the maximum correntropy Kalman filter (MCKF) can be utilized to provide robustness for the Kalman filter in the presence of non-Gaussian noise or large outliers~\cite{cinar2012hidden,izanlo,MCCNEW}. The MCKF uses the correntropy criterion instead of MMSE through which higher-order information of process and
measurement noises is used~\cite{liu2007correntropy,he2011maximum}.

Motivated by the aforementioned shortcomings of the existing controllers, that have been already designed for the influenza epidemic systems, and the desire to develop a new multi-objective controller for such systems, this work
is the first step towards designing a state estimation-based robust optimal controller
for influenza epidemics in an interactive human society (demonstrated in
Fig.~\ref{fig.FluFlo}) in the presence of modeling uncertainties and non-Gaussian noise.
The main contributions of this paper are as follows: (i) the state estimation of the influenza epidemics in an interactive human society;
(ii) the design of a robust optimal controller to minimize the population of susceptible and infected humans, while minimizing and bounding the rates of vaccination and antiviral treatment;
(iii) the proof of the UUB/convergence of tracking errors; and (iv) the robustness of the proposed algorithm in the presence of parameter perturbation and random entrance of individuals from the other societies.

In this paper, we begin by formulating an extended MCKF (EMCKF) algorithm to estimate the states of an influenza dynamical system while using the number of exposed and infected humans as measurement. With the aim of achieving the boundedness/convergence of system's errors with a minimal control effort, an online quadratic program (QP) is synthesized subject to a robust control Lyapunov function (RCLF). The joint QP-RCLF finds the optimal balance between control effort and stability of closed-loop system. The robust term is incorporated in the QP-RCLF framework to compensate for state estimation error and modeling uncertainties. The unified
state estimation-based controller QP-RCLF-EMCKF provides the convergence of susceptible and infected populations to a small neighborhood around the origin, while minimizing and bounding
the control effort. The UUB/convergence of tracking errors is finally proven using a Lyapunov stability argument. To assess the performance of the proposed
approach QP-RCLF-EMCKF, simulation results are carried out for the influenza epidemic model. Results show that the proposed controller successfully achieves the promised design specifications such as tracking and state estimation for this epidemiological system. Tests show that the QP-RCLF-EMCKF strategy provides appropriate robustness in the presence of parametric uncertainties and
random entrance of humans from other societies to the society of interest.

\indent The paper is organized as follows. \hyperref[section.sys-PS]{Section~\ref*{section.sys-PS}}
describes an influenza epidemic model and the problem statement.
\hyperref[section.est]{Section~\ref*{section.est}} presents
the state estimation framework using EMCKF algorithm.
\hyperref[section.cont]{Section~\ref*{section.cont}}
presents our proposed control strategy QP-RCLF-EMCKF.
\hyperref[section.Sim]{Section~\ref*{section.Sim}}
provides the simulation results.
\hyperref[section.Conc]{Section~\ref*{section.Conc}} presents
discussion, conclusion, and future work.

\section{Influenza Epidemic Model and Problem Statement} \label{section.sys-PS}

In this section, we begin by describing a dynamical model for the influenza epidemics and then present the problem statement. 

\subsection{Influenza epidemic model} \label{section.sys}

A state space representation of the influenza epidemics
can be described by the following nonlinear SEIAR model~\cite{Sharifi}:
\begin{align}\nonumber
\label{eq.model}
\dot{z}_{1}&=-\beta z_{1}\left(\epsilon z_{2}+(1-q)z_{3}+\delta z_{4} \right)-z_{1}u_{1}
\\ \nonumber
\dot{z}_{2}&=\beta z_{1}\left(\epsilon z_{2}+(1-q)z_{3}+\delta z_{4}\right)-\kappa z_{2}
 \\ \nonumber
\dot{z}_{3}&=p\kappa z_{2}-\alpha z_{3}-u_{2}z_{3}
\\ \nonumber
\dot{z}_{4}&=(1-p)\kappa z_{2}-\eta z_{4}
\\ 
\dot{z}_{5}&=\alpha\zeta z_{3}+z_{1}u_{1}+z_{3}u_{2}+\eta z_{4},
\end{align}
where $\underline{z}=[{z}_{1}, {z}_{2}, {z}_{3}, {z}_{4},{z}_{5}]^T=
[S, E, I, A, R]^T\in\Re^5$
denotes the state variables of the system with positive values; ${z}_{1}$ represents the population that is susceptible to get infected with influenza; ${z}_{2}$ is the
number of people who are infected with influenza but not yet infectious (exposed);
${z}_{3}$ stands for population that is infected and also infectious
with influenza symptoms; ${z}_{4}$ represents the number of individuals who are
influenza carriers but without any symptoms (asymptomatic); ${z}_{5}$ denotes the number of
recovered humans; $\underline{u}=[{u}_{1}, {u}_{2}]^T\in\Re^2$
is the vector of normalized control inputs
such that $0\leq{u}_{i}\leq1$ for $i=1,2$; ${u}_{1}$ is the rate of vaccination
of the susceptible population ${z}_{1}$; and ${u}_{2}$ is the rate of antiviral
treatment of the
infected population ${z}_{3}$. More details about this epidemiological model can be found
in~\cite{Inf1,Inf2,Sharifi}.

\subsection{Problem statement} \label{section.PS}

This paper aims to design a robust optimal controller to decrease the number of
susceptible $z_1$ and infected $z_3$ populations while using the minimum
possible rates of vaccination $u_1$ and antiviral treatment $u_2$. More importantly,
the normalized control inputs must be bounded between 0 and 1, which requires
a set of control constraints to be incorporated in the controller design.
For this purpose, an online QP control strategy is formulated by considering the RCLF and the above-mentioned input constraints
to generate a pointwise optimal control effort, while achieving the convergence of system's errors.

Since only $z_2$ and $z_3$ are measurable in practice,
the proposed controller uses the estimate of system's states (populations) as feedback in closed-loop system.
To achieve a robust state estimation of the influenza epidemics
in an interactive human society in
the presence of non-Gaussian noise, an EMCKF algorithm is employed and specifically developed 
for this dynamical system. A robust term
is also designed to robutify the system against state estimation error and parametric
uncertainties. The resulting state estimation-based control strategy QP-RCLF-EMCKF meets multiple
design objectives such as tracking, control optimality, state estimation,
and robustness. The UUB/convergence of all system solutions is proven
using a Lyapunov framework and the proposed controller is finally
validated by comprehensive simulation studies.

\section{State Estimation Using Extended Maximum Correntropy Kalman Filter (EMCKF)}
\label{section.est}

In this section, an EMCKF algorithm is described and presented to estimate the system states.
This filter only uses the number of exposed and infected humans
 ($z_2$ and $z_3$) as possible measurements. Consider the following
 general form of a nonlinear stochastic continuous-time system
 for the influenza epidemic model~\eqref{eq.model}
\begin{align}\label{KFsys}\nonumber
  \dot{z}&=f\left(z,u,\Theta,t\right)+w(t) \\ 
  y&=h\left(z,t\right)+v(t),
\end{align}
where $h\left(z,t\right)=[{z}_{2}, {z}_{3}]^T\in\Re^2$ is
the vector of measurable variables (populations) in the influenza epidemics; $w(t)\in\Re^5$ is the continuous-time
process noise vector of the system with covariance matrix $Q\in\Re^{5\times5}$;
$v(t)\in\Re^2$ is the continuous-time measurement noise with covariance $R\in\Re^{2\times2}$; and
$\Theta$ is the vector of actual system parameters as
\begin{align}\label{THE}
\Theta=[\beta, \epsilon, q, \delta, \kappa, p, \alpha, \eta, \zeta]^T\in\Re^9.
\end{align}

\begin{assumption} \label{th.as0}
The noises $w(t)$ and $v(t)$ are both uncorrelated,
Gaussian, and zero-mean. However, a shot noise
is enforced to the measurement noise $v(t)$ to model
the effects from the other societies on the main interactive society, which results in
a non-Gaussian noise as
\end{assumption}
\begin{align}\label{eq.noise}\nonumber
  w(t)&\sim N(0,Q) \\ 
  v(t)&\sim N(0,R)+\textrm{shot noise}.
\end{align}

\begin{assumption} \label{th.as1}
The nonlinear functions $f(.)\in\Re^5$ and $h(.)\in\Re^2$
are sufficiently smooth in $z$, such that they can be linearized using
the Taylor series expansions.
\end{assumption}

The EMCKF is similar to the EKF as they are based on linearization using
first-order Taylor series expansion.
Therefore, the following Jacobian matrices are used to linearize the system:
\begin{align}\label{jacobiansys}
  A=  \left. \frac{\partial f\left(z,u,\hat{\Theta},t\right)}{\partial z}
  \right|_{\hat{z}}\in\Re^{5\times5}, \ \
  C=  \left. \frac{\partial h}{\partial z}  \right|_{\hat{z}}\in\Re^{2\times5},
\end{align}
where $\hat{z}$ and $\hat{\Theta}$ are the estimations of $z$ and $\Theta$, respectively.
The initialization of the filter is given as:
\begin{align}\label{eq.init}\nonumber
\hat{z}(0)&=E\left[z(0)\right] \\ 
P(0)&=E[\left(z(0)-\hat{z}(0)\right) {\left(z(0)-\hat{z}(0)\right)}^T],
\end{align}
where $E(.)$ stands for the expected value operation; $P(0)$ is the
covariance of the initial estimate; and $z(0)$ and $\hat{z}(0)$ show the
initial value of the states and its estimates, respectively.

The state estimate and the EMCKF gain
for the continuous-time nonlinear system~\eqref{eq.model}
are formulated as follows~\cite{EstimSimon,fakoorian2016ground}:
\begin{align}\label{eq.sysstate-K}\nonumber
  \dot{\hat{z}}&=f(\hat{z},u,\hat{\Theta},t)+K\left[y(t)-h(\hat{z},t)\right] \\ 
  K(t)&=P(t)\nu(t)C^T (t)R^{-1} (t)
\end{align}
in which the time-varying gain $\nu(t)$ and the estimation error
covariance matrix $P(t)$ are defined as
\begin{align}\label{eq.correngain-Covup}\nonumber
\nu(t)=& G_\sigma \left(\|y(t) - C(t)\hat{z}(t)\|_{R(t)^{-1}}\right) \\ \nonumber
\dot{P}=&A(t)P(t)+ P(t)A^T(t) + Q(t)  \\
&-P(t)C^T (t)R^{-1}(t)C(t)P(t)
\end{align}
with the kernel function $G_{\sigma}(\| \cdot \|)$ defined as
\begin{align}\label{eq.ker}
G_{\sigma}(\| \cdot \|)=\textrm{exp}\left(\frac{-\| \cdot \|^2}{2\sigma^2}\right),
\end{align}
where $\|.\|$ stands for the Euclidean norm of a vector; $\|.\|_{R(t)^{-1}}$ denotes a weighted Euclidean norm of a vector (i.e., $\|x\|^2_{R(t)^{-1}}=x^TR(t)^{-1}x$ with $R(t)$\footnote{It should be pointed out that in this paper, the covariance matrices $R$ and $Q$ are considered to be constant and diagonal.} as a positive definite matrix); and $\sigma$ is the user-specified bandwidth (kernel size).

The EMCKF algorithm is robust against large outliers or non-Gaussian
noises, because when the system is perturbed by such noises, then
$\nu(t)\rightarrow 0 $ which prevents the divergence of the filter.
It can be seen that by picking a large value of $\sigma$, $\nu(t)\rightarrow 1$ and
the EMCKF reduces to the ordinary EKF.

\begin{assumption} \label{th.as2}
We assume that under the
EMCKF algorithm, the state estimation error $e_{e}=z-\hat{z}\in\Re^5$ is bounded.
\end{assumption}

\begin{remark} \label{th.rem11}
The EMCKF algorithm uses the estimate of system parameters $\hat{\Theta}$,
the measurements $(z_2, z_3)$, and the control signal $u$.
\end{remark}

The next section will formulate a state-estimation robust optimal control while utilizing the estimate of system's states provided by the EMCKF algorithm.

\section{Proposed Controller QP-RCLF-EMCKF} \label{section.cont}

With the estimate of the system's states from the previous section in hand,
this section is devoted to formulating the proposed controller
in order to minimize the susceptible and infected populations.
%ideally to zero (${z}_{1}\rightarrow 0$ and ${z}_{3}\rightarrow 0$).
Defining $z_{e}=[z_{1}, z_{3}]^T\in\Re^2$, the tracking objective
reduces to the convergence of $z_{e}$ to its desired minimum value $z^d_{e}$.
To achieve this objective, the first and third equations of Eq.~\eqref{eq.model} are taken
into account and can be written as follows
\begin{align}
\label{eq.modelcont}
\dot{z}_{e}=Y(z,\theta)-Z_{e} u
\end{align}
with
\begin{align}\label{eq.modelcont2}
Y(z,\theta)=\left[ \begin{array}{c}
              \theta^T_1\Phi_1(z)  \\ \theta^T_2\Phi_2(z)
            \end{array} \right],
\end{align}
where the basis functions $\Phi_1(z)$ and $\Phi_2(z)$, the parameter vectors $\theta_1$ and $\theta_2$, and the control map $Z_{e}$ including positive diagonal elements are defined as
\begin{align}\label{eq.MAP}\nonumber
\Phi_1(z)=&[-z_{1}z_{2}, -z_{1}z_{3}, -z_{1}z_{4}]^T\in\Re^3
\\ \nonumber
\Phi_2(z)=&[z_{2}, -z_{3}]^T\in\Re^2
\\ \nonumber
\theta_1=&[\epsilon\beta, \beta(1-q), \beta\delta]^T\in\Re^3
\\ \nonumber
\theta_2=&[p\kappa,\alpha]^T\in\Re^2
\\ 
Z_{e}=&\textrm{diag}(z_{1}, z_{3})\in\Re^{2\times2}.
\end{align}

Let us define $e=\hat{z}_e-z^d_e$ as the tracking error vector.
Defining $\hat{z}_{e}=[\hat{z}_1,\hat{z}_3]^T$ and $e_{e_{1,3}}=[e_{e_{1}},e_{e_{3}}]^T=z_{e}-\hat{z}_{e}$,
the tracking error can be redefined as
\begin{align}\label{eq.ertr}
e=z_e-e_{e_{1,3}}-z^d_e.
\end{align}

\begin{assumption} \label{th.as4}
	Assume that the desired value $z^d_e$ is bounded and of class $\mathcal{C}^1$ (i.e., $z^d_e$ is continuously differentiable)\footnote{A function is said to be of class $\mathcal{C}^n$ if its first $n$ derivatives all exist and are continuous.}.
\end{assumption}

Using Eqs.~\eqref{eq.modelcont} and~\eqref{eq.ertr},
the error dynamics are obtained as
\begin{align}\label{eq.err}
\dot{e}=Y-Z_{e} u-\dot{z}^d_e-\dot{e}_{e_{1,3}}.
\end{align}

Using the notion of the feedback linearization,
assuming that $\dot{e}_{e_{1,3}}=0$, and picking the following feedback control law
\begin{align}\label{eq.lincont}
u=Z_{e}^{-1}(Y-\mu-\dot{z}^d_e),
\end{align}
the error dynamics~\eqref{eq.err} are transferred
to the linear system $\dot{e}=\mu$ with $\mu$ as the virtual input vector.

However, it should be pointed out that (i) the vector $\dot{e}_{e_{1,3}}$ is nonzero, (ii) the actual system parameters $\theta$ are not
perfectly known, and (iii) the accurate measurement of state variables $z$ is not available
to the controller. To include the estimated state $\hat{z}$ and 
parameters $(\hat{\theta}_1,\hat{\theta}_2)$ (Items ii and iii), the feedback law~\eqref{eq.lincont}
is modified  as
\begin{align}\label{eq.lincont2}
u=\hat{Z}_{e}^{-1}(\hat{Y}-\mu-\dot{z}^d_e),
\end{align}
where
\begin{align}\label{eq.Yhat}\nonumber
\hat{Y}&=[\hat{Y}_1,\hat{Y}_2]^T=\left[ \begin{array}{c}
              \hat{\theta}^T_1\hat{\Phi}_1  \\ \hat{\theta}^T_2\hat{\Phi}_2
            \end{array} \right], \ \
        \hat{Z}_{e}=\textrm{diag}(\hat{z}_{1}, \hat{z}_{3}),
        \\ \nonumber
       \hat{\Phi}_1&=[-\hat{z}_{1}\hat{z}_{2}, -\hat{z}_{1}\hat{z}_{3},
       -\hat{z}_{1}\hat{z}_{4}]^T, \
       \hat{\Phi}_2=[\hat{z}_{2}, -\hat{z}_{3}]^T,
       \\ 
       \hat{\theta}_1&=[\hat{\epsilon}\hat{\beta}, \hat{\beta}(1-\hat{q}),
       \hat{\beta}\hat{\delta}]^T, \ \ \ \ \ \ \ \
       \hat{\theta}_2=[\hat{p}\hat{\kappa},\hat{\alpha}]^T.
\end{align}

Substituting the control law~\eqref{eq.lincont2} into
the error dynamics~\eqref{eq.err} in the presence of a nonzero
$\dot{e}_{e_{1,3}}$ (Item i), one has
\begin{align}\label{eq.err2}
\dot{e}=Y-Z_{e}\hat{Z}_{e}^{-1}\left(\hat{Y}-\mu-\dot{z}^d_e\right)-\dot{z}^d_e-\dot{e}_{e_{1,3}}.
\end{align}

By rewriting the control map as
\begin{align}\label{eq.Rmap}
Z_{e}=\textrm{diag}(\hat{z}_{1}+e_{e_1}, \hat{z}_{3}+e_{e_3}),
\end{align}
the term $Z_{e}\hat{Z}_{e}^{-1}$ can be stated as
\begin{align}\label{eq.zzhat}
Z_{e}\hat{Z}_{e}^{-1}=I+\textrm{diag}(\frac{e_{e_1}}{\hat{z}_{1}},\frac{e_{e_3}}{\hat{z}_{3}}),
\end{align}
where $e_{e_1}=z_{1}-\hat{z}_{1}$ and $e_{e_3}=z_{3}-\hat{z}_{3}$.
Then, by defining $\Delta\Phi_i=\Phi_i-\hat{\Phi}_i$
and $\Delta\theta_i=\theta_i-\hat{\theta}_i$ for $i=1,2$,
the vector $Y$ can be expressed as
 \begin{align}\label{eq.yyhat}
 Y=\left[ \begin{array}{c}
              (\hat{\theta}^T_1+\Delta\theta^T_1)(\hat{\Phi}_1+\Delta\Phi_1)
              \\ (\hat{\theta}^T_2+\Delta\theta^T_2)(\hat{\Phi}_2+\Delta\Phi_2)
            \end{array} \right]
            =\hat{Y}+\Delta_1
\end{align}
in which 
%$\Delta_1(e_{est},e,\hat{\theta},\hat{z},\Delta\theta)\in\Re^2$
$\Delta_1\in\Re^2$
is defined as
  \begin{align}\label{eq.delt1}
\Delta_1=\left[ \begin{array}{c}
\hat{\theta}^T_1\Delta\Phi_1+\Delta\theta^T_1\hat{\Phi}_1+\Delta\theta^T_1\Delta\Phi_1
\\ \hat{\theta}^T_2\Delta\Phi_2+\Delta\theta^T_2\hat{\Phi}_2+\Delta\theta^T_2\Delta\Phi_2
            \end{array} \right].
\end{align}

Now, substituting Eqs.~\eqref{eq.zzhat} and~\eqref{eq.yyhat} into Eq.~\eqref{eq.err2} yields
\begin{align}\label{eq.err3}
\dot{e}=\mu+\Delta
\end{align}
for which the uncertainty term $\Delta\in\Re^2$ is described as 
\begin{align}\label{eq.err3N}
\Delta=\Delta_1+\Delta_2-\dot{e}_{e_{1,3}},
\end{align}
where
%$\Delta_2(e_{est},e,\hat{\theta},\hat{z},\mu,\dot{z}^d_e)\in\Re^2$ 
$\Delta_2\in\Re^2$ is
  \begin{align}\label{eq.delt2}
\Delta_2=-\textrm{diag}(\frac{e_{e_1}}{\hat{z}_{1}},\frac{e_{e_3}}{\hat{z}_{3}})
\left(\hat{Y}-\mu-\dot{z}^d_e\right).
\end{align}

In the next section, to provide the context for the uncertainty term $\Delta$, its properties will be studied in detail.

\subsection{Properties of the uncertainty term $\Delta$} \label{section.prop}
Throughout this section, we rely on the following property. 

\begin{prop} \label{th.prop1}
Let us define the whole population of the society as $N=\sum_{i=1}^5 z_i$ whose variation can be obtained by the summation of all compartmental dynamics presented in~\eqref{eq.model}  
\begin{align}\label{eq.NEWN}
\dot{N}=-\alpha(1-\zeta)z_3,
\end{align}
where $\alpha>0$ denotes the recovery rate for the symptomatic infected people and $0<\zeta<<1$ is the fatality rate of the influenza. In view of~\eqref{eq.NEWN}, it follows that the whole population $N$ is a decaying upper bounded time-varying function such that $N(t)\leq N_0$, where $N_0>0$ is its initial magnitude. Hence, all compartmental variables $z_i$ for $i=1,\dots,5$ remain bounded during the treatment time such that $z_i\leq N(t)\leq N_0$.
Whereby, according to Assumption~\ref{th.as2}, the estimates of all system variables $z_i$ are also bounded.
\end{prop}

In the following, we begin by expanding each of the components in~\eqref{eq.err3N} and then describe the uncertainty term $\Delta$ as a linear function of $\|e\|$ plus a bounded term. 

\medskip
\subsubsection{Term $\Delta_1$} \label{section.D1}

Utilizing the definitions of the tracking and estimation errors from Assumption~\ref{th.as2} and Eq.~\eqref{eq.ertr}, the vectors $\Delta \Phi_1$, $\Delta \Phi_2$, and $\hat{\Phi}_1$ can be written as
\begin{equation*}\label{eq.NEW11}
\Delta \Phi_1=\left[ \begin{array}{c}
              -z_1z_2+\hat{z}_1\hat{z}_2  \\ -z_1z_3+\hat{z}_1\hat{z}_3 \\ -z_1z_4+\hat{z}_1\hat{z}_4
            \end{array} \right] 
            =A_1e+W_1,
\end{equation*}
\begin{equation}\label{eq.NEW12}
\Delta \Phi_2=\left[ \begin{array}{c}
              e_{e_2}  \\ -e_{e_3}
            \end{array} \right], \ \ \ \hat{\Phi}_1=A_2e+W_2
\end{equation}
with
\begin{equation*}\label{eq.NEW13}
A_1=-\left[ \begin{array}{cc}
              e_{e_2} & 0  \\ e_{e_3} & e_{e_1} \\ e_{e_4} & 0
            \end{array} \right], \ \ \ 
A_2=-\left[ \begin{array}{cc}
              \hat{z}_2 & 0  \\ \hat{z}_4 & 0 \\ 0 & \hat{z}_1
            \end{array} \right]            
\end{equation*}
\begin{equation}\label{eq.NEW14}
W_1=-\left[ \begin{array}{c}
              e_{e_2}z^d_{e_1}+e_{e_1}z_2  \\ e_{e_1}e_{e_3}+e_{e_1}z^d_{e_2}+e_{e_3}z^d_{e_1} \\ e_{e_4}z^d_{e_1}+e_{e_1}z_4
            \end{array} \right] \ \
            W_2=-\left[ \begin{array}{c}
              \hat{z}_2z^d_{e_1}  \\ \hat{z}_4z^d_{e_1} \\ \hat{z}_1z^d_{e_2}
            \end{array} \right].
\end{equation}

In view of~\eqref{eq.NEW12} and~\eqref{eq.NEW14}, the term $\Delta_1$ has the alternative form 
  \begin{equation}\label{eq.NEW16}
\Delta_1=A_3e+W_3
\end{equation}
where
\begin{equation}\label{eq.NEW17}\nonumber
A_3=\left[ \begin{array}{c}
              \theta^T_1 A_1+\Delta \theta^T_1 A_2  \\ 0
            \end{array} \right], \\ 
 W_3=\left[ \begin{array}{c}
              \theta^T_1 W_1+\Delta \theta^T_1 W_2  \\ \theta^T_2 \Delta\Phi_1+ \Delta\theta^T_2 \hat{\Phi}_1
            \end{array} \right].           
\end{equation}

According to Assumptions~\ref{th.as2} and~\ref{th.as4}, and Property~\ref{th.prop1}, all terms in the matrices $A_1$, $A_2$, and $W_1$ and the vectors $W_2$, $\Delta \Phi_2$, and $\hat{\Phi}_2$ are bounded. This coupled with the boundedness of the vectors $\theta_i$, $\hat{\theta}_i$, and $\Delta \theta_i$ for $i=1,2$ concludes that the term $\Delta_1$ is bounded by a linear function of $\|e\|$ plus a bounded term $W_3$ such that
  \begin{align}\label{eq.NEW18}
\|\Delta_1\|\leq\bar{A}_3\|e\|+\bar{W}_3,
\end{align}
where $\bar{A}_3$ and $\bar{W}_3$ are positive scalars such that $\|A_3\|\leq\bar{A}_3$ and $\|W_3\|\leq\bar{W}_3$.

\medskip
\subsubsection{Term $\Delta_2$} \label{section.D2}

In view of~\eqref{eq.lincont2}, one obtains $\mu=\hat{Y}-\dot{z}^d_e-\hat{Z}_{e}u$ using which the term $\Delta_2$ reduces to 
\begin{align}\label{eq.NEW19}
\Delta_2=-W_4u
\end{align}
with 
\begin{align}\label{eq.NEW19}
W_4=\left[ \begin{array}{cc}
              e_{e_1} & 0\\ 0 & e_{e_3}
            \end{array} \right].
\end{align}            
            
In \hyperref[section.contqp]{Section~\ref*{section.contqp}}, we will synthesize a QP optimization problem through which the control input $u_i$ for $i=1,2$ is enforced to always stay between 0 and 1, i.e., $\|u\|\leq u_{0}$ with a positive scalar $u_{0}$. This bounding of the control signal along with the boundedness of $e_{e_1}$ and $e_{e_3}$ implies that 
  \begin{align}\label{eq.NEW20}
\|\Delta_2\|\leq\bar{W}_4u_{0},
\end{align}
where $\|W_4\|\leq\bar{W}_4$ with $\bar{W}_4>0$.

\medskip
\subsubsection{Term $\dot{e}_{e_{1,3}}$} \label{section.eedot}

In view of Eq.~\eqref{eq.sysstate-K}, the derivative of the estimation error for the number of susceptible and infected populations is
  \begin{align}\label{eq.NEW21}\nonumber
\dot{e}_{e_{1,3}}=&\dot{z}_e-\dot{\hat{z}}_e \\ 
=&Y-Z_{e} u-\hat{Y}+\hat{Z}_{e}u-K_{1,3}(y-\hat{y}),
\end{align}
where $K_{1,3}\in\Re^{2\times2}$ is a matrix whose rows represent the first and third rows of the Kalman gain. Utilizing the definitions $e_{e_i}=z_{i}-\hat{z}_{i}$ for $i=1,\dots,5$, and $\Delta\Phi_i=\Phi_i-\hat{\Phi}_i$
and $\Delta\theta_i=\theta_i-\hat{\theta}_i$ for $i=1,2$, one has
  \begin{align}\label{eq.NEW22}
\dot{e}_{e_{1,3}}=\left[ \begin{array}{c}
              \theta^T_1 \Delta\Phi_1+ \Delta\theta^T_1 \hat{\Phi}_1\\ 
              \theta^T_2 \Delta\Phi_2+ \Delta\theta^T_2 \hat{\Phi}_2
            \end{array} \right]-
            W_4u+W_5
\end{align}
with $W_5=-K_{1,3} \ [e_{e_2},e_{e_3}]^T$.
A careful inspection of Eq.~\eqref{eq.NEW22} reveals that the first term is equal to the term $\Delta_1$ and therefore, one can write
  \begin{align}\label{eq.NEW23}
\dot{e}_{e_{1,3}}=A_3e+W_3-W_4u+W_5
\end{align}
in which since $A_3$, $W_3$, $W_4$, $W_4$, $K_{1,3}$, and $u$ are all bounded, the bound for $\|\dot{e}_{e_{1,3}}\|$ is obtained as  
  \begin{align}\label{eq.NEW24}
\|\dot{e}_{e_{1,3}}\|\leq\bar{A}_3\|e\|+\bar{W}_3+\bar{W}_4u_{0}+\bar{W}_5,
\end{align}
where $\bar{W}_5$ is a positive scalar such that $\|W_5\|\leq\bar{W}_5$.

\medskip

Using the previously computed bounds, the uncertainty term $\Delta$ can be stated as a linear function of $\|e\|$ plus a bounded term
  \begin{align}\label{eq.NEW24} \nonumber
\|\Delta\|\leq& \|\Delta_1\|+\|\Delta_2\|+\|\dot{e}_{e_{1,3}}\| \\ 
\nonumber
\leq& \bar{A}_3\|e\|+\bar{W}_3+\bar{W}_4u_{0}+\bar{A}_3\|e\|+\bar{W}_3+\bar{W}_4u_{0}+\bar{W}_5 \\ 
=&\underbrace{2\bar{A}_3}_{\bar{A}}\|e\|+\underbrace{2\left(\bar{W}_3+\bar{W}_4u_{0}\right)+\bar{W}_5}_{\bar{W}},
\end{align}
where $\bar{A}$ and $\bar{W}$ are two positive scalars.

Employing the proposed feedback control law~\eqref{eq.lincont2},
the error dynamics~\eqref{eq.err} are partially linearized as presented in
Eq.~\eqref{eq.err3}.
Then, the problem reduces to designing the virtual input $\mu$ to guarantee the UUB/convergence of error trajectory $e$ while compensating
for the uncertainty $\Delta$. For this purpose, the next subsection will present a RCLF to ensure boundedness/convergence of the tracking error in a pointwise optimal fashion.   

%\begin{remark}\label{th.rem2}
%It should be noted that the uncertainty $\Delta$ is a general function of
%the state estimation error, tracking error, estimated parameters,
%parameter estimation error,
%and control input as
%\begin{align}\label{eq.DeltaG}
%\Delta=g(e_{est},e,\hat{\theta},\Delta\theta,\hat{z},\mu,\dot{z}^d_e,\dot{e}_{est},t).
%\end{align}
%\end{remark}

%\begin{assumption} \label{th.as6}
%For any bounded control input $\mu = \mu(\hat{z},\hat{\theta}, t)$, there exists
%a positive bound 
%%$D(e_{est},e,\Delta\theta,\hat{z},\mu,t)$
%$D$
% for the uncertainty $\Delta$.
%\end{assumption}

\subsection{Robust control Lyapunov function (RCLF)} \label{section.contmin}

In this section, we begin by considering the special case
of $\Delta=0$ based on which the system~\eqref{eq.err3} reduces to 
\begin{equation}\label{eq.err3A}
\dot{e}=\mu.
\end{equation}

A function $V(e)$ is an exponentially stabilizing
control Lyanpunov function (ES-CLF) for the system~\eqref{eq.err3A}, if the following conditions are met~\cite{Ames2014}:
\begin{align} \label{eq.CLF}
  a_{1} \|e\|^2 &\leq V(e)\leq a_{2} \|e\|^2 \\
  \label{eq.CLF2}
  \dot{V}(e)&\leq -\lambda V(e),
\end{align}
where $a_{1}, a_{2}, \lambda>0$. A candidate ES-CLF for
the system~\eqref{eq.err3A} is then suggested as
 \begin{align}
\label{eq.Ly}
V(e)=\frac{1}{2}e^T e
\end{align}
whose time derivative is
 \begin{align}
\label{eq.DLy}
\dot{V}(e)=e^T\dot{e}=e^T\mu.
\end{align}

Now, by choosing $\mu=-\lambda e$ and based on Eq.~\eqref{eq.CLF2}, $V$ is ES-CLF.
As an alternative, $\dot{V}(e)$ in Eq.~\eqref{eq.DLy} can
be expressed in terms of the main control input $u$. 

For this purpose,
substituting the virtual input $\mu$ from Eq.~\eqref{eq.lincont2} into $\dot{V}(e)$ yields
  \begin{align}
\label{eq.LyD}
\dot{V}(e)=L_{f}V(e)+L_{g}V(e)u
\end{align}
with $L_{f}V(e)\in\Re$ and $L_{g}^T V(e)\in\Re^2$ as
 \begin{align}\nonumber
\label{eq.Lie}
L_{f}V(e)&=e^T(\hat{Y}-\dot{z}^d_e) \\ 
L_{g}V(e)&=-e^T\hat{Z}_{e}
\end{align}
based on which $\phi_{0}\in\Re$ and $\phi_{1}\in\Re^2$ are defined as 
\begin{align}\nonumber
\label{eq.phi}
\phi_{0}(e)&=L_{f}V(e)+\lambda V(e)\\ 
\phi_{1}(e)&=L_{g}^TV(e).
\end{align}

Then, substituting Eq.~\eqref{eq.LyD} into Eq.~\eqref{eq.CLF2},
and using the definitions of $\phi_{0}$ and $\phi_{1}$ from Eq.~\eqref{eq.phi},
the inequality constraint~\eqref{eq.CLF2} can be expressed as
\begin{equation}
\label{eq.Expphi} 
\phi_{0}+\phi_{1}^Tu \leq 0,
\end{equation}
which is called the \textit{CLF constraint}. 

Now, a family of controllers that can minimize the control input
$u$ w.r.t. the inequality constraint~\eqref{eq.Expphi} can be
defined using the following pointwise min-norm control (PWMC) law~\cite{Freeman}:
\begin{equation} \label{eq.PWMC}
  u(\phi_{0},\phi_{1})=
    \begin{cases}
      -\frac{\phi_{0}(e)\phi_{1}(e)}
            {\phi_{1}^T(e)\phi_{1}(e)}
     & \text{if}\ \phi_{0}(e)>0 \\
       0
     & \text{if}\ \phi_{0}(e)\leq0
    \end{cases}.
\end{equation}

However, this control law can only guarantee the exponential
convergence of $e$ to zero in the absence of the quantity $\Delta$. We now consider the general case in which $\Delta\neq0$ for the error dynamics~\eqref{eq.err3}. 

\begin{theorem}\label{th.Main1}
Consider the Lyapunov function~\eqref{eq.Ly} and the control law~\eqref{eq.PWMC}. Under the Assumptions~\ref{th.as0}, ~\ref{th.as1},~\ref{th.as2}, and~\ref{th.as4} and Property~\ref{th.prop1}, the tracking error norm remains less than $B_r=2\bar{W}/\Lambda$ with $\Lambda=\lambda-2\bar{A}>0$ at all time for all $\Theta\in\Re^9$, any $e(0)\in\Re^2$, and any bounded $e_e(0)\in\Re^5$. 
\end{theorem}

\begin{proof}
In the presence of the uncertainty $\Delta$,
$\dot{V}(e)$ in Eq.~\eqref{eq.LyD} converts to
\begin{align}
\label{eq.LyD2}
\dot{V}(e)=e^T(\mu+\Delta)=L_{f}V(e)+L_{g}V(e)u+e^T\Delta.
\end{align}

By substituting the PWMC law~\eqref{eq.PWMC}
into Eq.~\eqref{eq.LyD2} when $\phi_{0}(e)>0$, one can write
\begin{align}
\label{eq.LyD3}
\dot{V}(e)=-\frac{\lambda}{2}e^Te+e^T\Delta
\end{align}
which implies that
\begin{align}
\label{eq.LyD4}
\dot{V}(e)\leq-\frac{\lambda}{2}\|e\|^2+\|e\|\|\Delta\|.
\end{align}

Substitute the calculated bound for $\|\Delta\|$ from \hyperref[section.prop]{Section~\ref*{section.prop}} to have
\begin{align}\nonumber
\label{eq.LyD5}
\dot{V}(e)\leq&-\frac{\lambda}{2}\|e\|^2+\|e\|\left(\bar{A}\|e\|+\bar{W}\right)
\\ 
=&-\frac{1}{2}(\lambda-2\bar{A})\|e\|^2+\bar{W}\|e\|.
\end{align}
By defining $\Lambda=\lambda-2\bar{A}>0$ with $\lambda>2\bar{A}$, $\dot{V}(e)<0$ outside the set
\begin{align}
\label{eq.compar}
\mathcal{S}_0=\{e: \ \  \|e\|\leq\frac{2\bar{W}}{\Lambda}=B_r\}.
\end{align}

This implies that the tracking error norm remains less than $B_r$ at all time when $\phi_{0}(e)>0$. 
In case that $\phi_{0}(e)\leq0$, we have $u=0$ for which Eq.~\eqref{eq.LyD2} becomes
\begin{align}
\label{eq.LyD2c}
\dot{V}(e)=L_{f}V(e)+e^T\Delta.
\end{align}

On the other hand, $\phi_{0}(e)\leq0$ implies that
\begin{align}
\label{eq.cond1}
L_{f}V(e)\leq -\lambda V(e)=-\frac{\lambda}{2}e^Te.
\end{align}
Using Eq.~\eqref{eq.LyD2c} and Eq.~\eqref{eq.cond1} and following the
same steps as in Eq.~\eqref{eq.LyD5}, we conclude that $e$ is bounded in the same ball $B_r$ as in Eq.~\eqref{eq.compar}.

The analysis can be further extended to show the exponential convergence of the tracking error vector to the set $\mathcal{S}_0$. For this purpose, apply the Young’s inequality for~\eqref{eq.LyD5} on the term $\bar{W}\|e\|$ to obtain
\begin{align}
\label{eq.QWN1}
\dot{V}(e)\leq-\frac{\bar{\Lambda}}{2}\|e\|^2+\frac{\bar{W}^2}{2}\leq-\bar{\Lambda}V(e)+\frac{\bar{W}^2}{2},
\end{align}
where $\bar{\Lambda}=\Lambda-1>0$ with $\Lambda>1$. Applying the Comparison lemma~\cite{khalil} (Lemma 3.4), one obtains
\begin{align}
\label{eq.QWN2}
V(e)\leq e^{-\bar{\Lambda} t}V(0)+\frac{\bar{W}^2}{2\bar{\Lambda}}.
\end{align}

This implies that $V$ exponentially converges to a ball of size $\bar{W}^2/(2\bar{\Lambda})$ with
exponential converge rate $\bar{\Lambda}$. Hence, since $\|e\|\leq \sqrt{2V(e)}$, the tracking error $e$ will exponentially converge to the small compact set $\mathcal{S}_0$.  
\end{proof}

\begin{remark} \label{th.remn1}
The size of the convergence ball $B_r$ is determined by the parameter $\Lambda$ and the bound $\bar{W}$, where the former can be tuned by users and the latter depends on the parameter uncertainties and the state estimation error.      
\end{remark}

\begin{remark} \label{th.rem1}
The error trajectory $e$
converges to a smaller ball for
smaller state estimation error and parameter estimation error (smaller $\bar{W}$).
The effect of the uncertainty $\Delta$ can be also mitigated
by choosing a sufficiently large value of $\lambda$.
However, this may cause higher control effort and unpleasant system solutions.
\end{remark}

It is seen that the PWMN control law~\eqref{eq.PWMC}
with defined $\phi_{0}$ provides the boundedness of $e$
in a compact ball with size $B_r$.
With the aim of compensating the uncertainty term $\Delta$
and reducing the size of the ultimate ball
without manipulating the convergence rate,
the robust term 
\begin{align}
\label{eq.term}
C_{rob}=K_r \|e\|, \ \ K_r>0
\end{align}
 is incorporated into $\phi_{0}$ to obtain
\begin{align}
\label{eq.phir}
\phi_{0_{rob}}&=L_{f}V(e)+\lambda V(e)+C_{rob}. \\ \nonumber
\end{align}

Employing Eq.~\eqref{eq.phir}, the inequality constraint~\eqref{eq.Expphi} can be rewritten as
\begin{align}\label{eq.RCLF} 
\phi_{0_{rob}}+\phi_{1}^T u \leq 0
\end{align}
which is called the \textit{RCLF constraint}.

So now, the modified control law based upon $\phi_{0_{rob}}$ is suggested as
\begin{equation} \label{eq.PWMCNEW}
u(\phi_{0_{rob}},\phi_{1})=
    \begin{cases}
      -\frac{\phi_{0_{rob}}(e)\phi_{1}(e)}
            {\phi_{1}^T(e)\phi_{1}(e)}
     & \text{if}\ \phi_{0_{rob}}(e)>0 \\
       0
     & \text{if}\ \phi_{0_{rob}}(e)\leq0
    \end{cases}.
\end{equation}

\begin{theorem}\label{th.Main2}
Consider the Lyapunov function~\eqref{eq.Ly}, the robust component~\eqref{eq.term}, and the control law~\eqref{eq.PWMCNEW}. Under the Assumptions~\ref{th.as0}, ~\ref{th.as1},~\ref{th.as2}, and~\ref{th.as4} and Property~\ref{th.prop1}, if $K_r<\bar{W}$, then $\|e\|$ remains less than $B_{r_{rob}}=2(\bar{W}-K_r)/\Lambda$ at all time for all $\Theta\in\Re^9$, any $e(0)\in\Re^2$, and any bounded $e_e(0)\in\Re^5$. The convergence of $e$ to the compact ball $B_{r_{rob}}$ is globally exponential. However if $K_r\geq \bar{W}$, then $e$
asymptotically converges to zero as $t\rightarrow \infty$. 
\end{theorem}
\begin{proof}

Utilizing the control law~\eqref{eq.PWMCNEW} in case that $\phi_{0}(e)>0$, $\dot{V}(e)$ of Eq.~\eqref{eq.LyD3} can be written as
\begin{align}
\label{eq.LyD3N}
\dot{V}(e)=-\frac{\lambda}{2}e^Te+\Delta e^T-K_r\|e\|.
\end{align}
Hence,
\begin{align}
\label{eq.LyD4N}
\dot{V}(e)\leq-\frac{\lambda}{2}\|e\|^2+\|\Delta\|\|e\|-K_r\|e\|.
\end{align}

Once again, using the calculated bound of $\|\Delta\|$ from \hyperref[section.prop]{Section~\ref*{section.prop}}, one has
\begin{align}
\label{eq.LyD4N}
\dot{V}(e)\leq-\frac{1}{2}\Lambda\|e\|^2+(\bar{W}-K_r)\|e\|.
\end{align}

Here, two cases can be considered on selecting the robust gain $K_r$:

\medskip
\begin{case}[$K_r<\bar{W}$: uniform ultimate boundedness]
	
In this case, $\dot{V}(e)<0$ outside the set
\begin{align}
\label{eq.compar2}
\mathcal{S}_1=\{e: \ \  \|e\|\leq\frac{2(\bar{W}-K_r)}{\Lambda}=B_{r_{rob}}\},
\end{align}
which follows that the size of the new convergence ball is $2(\bar{W}-K_r)/\Lambda$.
This implies that employing the robust term $C_{rob}$ with a positive gain that satisfies $K_r<\bar{W}$ reduces
the size of the ultimate bound on the tracking error $e$. In this case, the size of $B_{r_{rob}}$ is determined by the parameter $\Lambda$ and the discrepancy between the gain $K_r$ and the bound $\bar{W}$.

Once again, to ensure that the convergence of $e$ to the set $\mathcal{S}_1$ is exponential, we apply the Young’s inequality for~\eqref{eq.LyD4N} on the term $(\bar{W}-K_r)\|e\|$ to have
\begin{align}
\label{eq.QWN3}
\dot{V}(e)\leq-\frac{\bar{\Lambda}}{2}\|e\|^2+\frac{(\bar{W}-K_r)^2}{2}\leq-\bar{\Lambda}V(e)+\frac{(\bar{W}-K_r)^2}{2},
\end{align}
for which applying the Comparison lemma yields
\begin{align}
\label{eq.QWN4}
V(e)\leq e^{-\bar{\Lambda} t}V(0)+\frac{(\bar{W}-K_r)^2}{2\bar{\Lambda}}.
\end{align}

This concludes exponential convergence of $V$ to a small neighborhood around the origin for which the size of the neighborhood is $(\bar{W}-K_r)^2/(2\bar{\Lambda})$ and the
exponential convergence rate is $\bar{\Lambda}$. This coupled with the the radial unboundedness of the Lyapunov function $V$ follows that the convergence of $e$ to the set $\mathcal{S}_1$ is globally 
exponential.

\end{case}

\begin{table*}[t]
	\centering
	\caption{Parameters of the nonlinear SEIAR model~\eqref{eq.model}~\cite{Sharifi}}
	\label{table.syspara2}
	\centering
	\begin{tabular}{ccc}
		\hline
		Parameter & Description & Values\\
		\hline
		\hline
		\centering
		$\kappa$ & Transition rate for the exposed & 0.526 \\
		$\alpha$ & Recovery rate for the infected & 0.244\\
		$\eta$ & Recovery rate for the asymptomatic & 0.244\\
		p & Fraction of developing symptoms & 0.667\\
		$\zeta$ & Fatality rate  & 0.98\\
		$\epsilon$& Infectivity reduction factor for the exposed & 0\\
		$\delta$ & Infectivity reduction factor for the asymptomatic & 1\\
		q & Contact reduction by isolation & 0.5
	\end{tabular}
\end{table*}

\medskip
\begin{case}[$K_r\geq \bar{W}$: asymptotic convergence]
	
In this case, picking a sufficiently large robust gain in such a way that $K_r=\bar{W}+\kappa_r$ with $\kappa_r>0$ results in
\begin{align}
\label{eq.LyD5N}
\dot{V}(e)\leq-\frac{\lambda}{2}\|e\|^2-\kappa_r\|e\|.
\end{align}

This concludes that $\dot{V}$ becomes negative definite, which implies that $e$
asymptotically converges to zero as $t\rightarrow \infty$.
\end{case}
\end{proof}

\begin{remark} \label{th.rem2}
Although the larger robust gain $K_r$ provides
better tracking performance, it results in a higher control signal
($K_r$ directly contributes to the control law $u$).
On the other hand, the smaller $K_r$ provides a better
control optimality, while the tracking error possesses a larger ultimate bound.
Thus, a trade off should be made between control optimality and tracking
performance when choosing the robust gain $K_r$.
\end{remark}

\begin{remark} \label{th.rem3}
The proposed control strategy with the RCLF structure renders stronger conclusion
for the stability of closed-loop system in the presence of uncertainty $\Delta$.
\end{remark}

\begin{table}[t]
	\centering
	\caption{Design parameters of the proposed QP-RCLF-EMCKF}
	\label{table.syspara}
	\centering
	\begin{tabular}{c|cccc}
		\hline
		& Parameter & Value & Location\\
		\hline
		\hline
		\centering
		\multirow{4}{1.5cm}{Filter} & P(0) & 1$I_5$ & Eq.~\eqref{eq.init}\\
		& R & 0.01$I_2$ & Eq.~\eqref{eq.sysstate-K}\\
		& Q & 1$I_5$ & Eq.~\eqref{eq.correngain-Covup}\\
		& $\sigma$ & 0.01 & Eq.~\eqref{eq.ker}\\
		\hline
		\centering
		\multirow{5}{1.5cm}{Controller}
		& $\lambda$ & 1 & Eq.~\eqref{eq.phir}\\
		& $K$ & 2 & Eq.~\eqref{eq.phir}\\
		& $c$ & 10 & Eq.~\eqref{eq.HB}\\
		& $\bar{u}_1$, $\bar{u}_2$ & 1 & Eq.~\eqref{eq.AB}\\
		& $\underline{u}_1$, $\underline{u}_2$ & 0 & Eq.~\eqref{eq.AB}\\
	\end{tabular}
\end{table}

With the formulation of the RCLF in hand, the next subsection will unify the EMCKF and the RCLF through synthesizing a QP optimization framework.   

\subsection{Unified controller QP-RCLF-EMCKF} \label{section.contqp}
\begin{figure}[t]
	\vspace{-2em}
	\centering
	\begin{subfigure}[b]{0.17\textwidth}
		\hspace*{-4em}
		\includegraphics[scale=0.4]{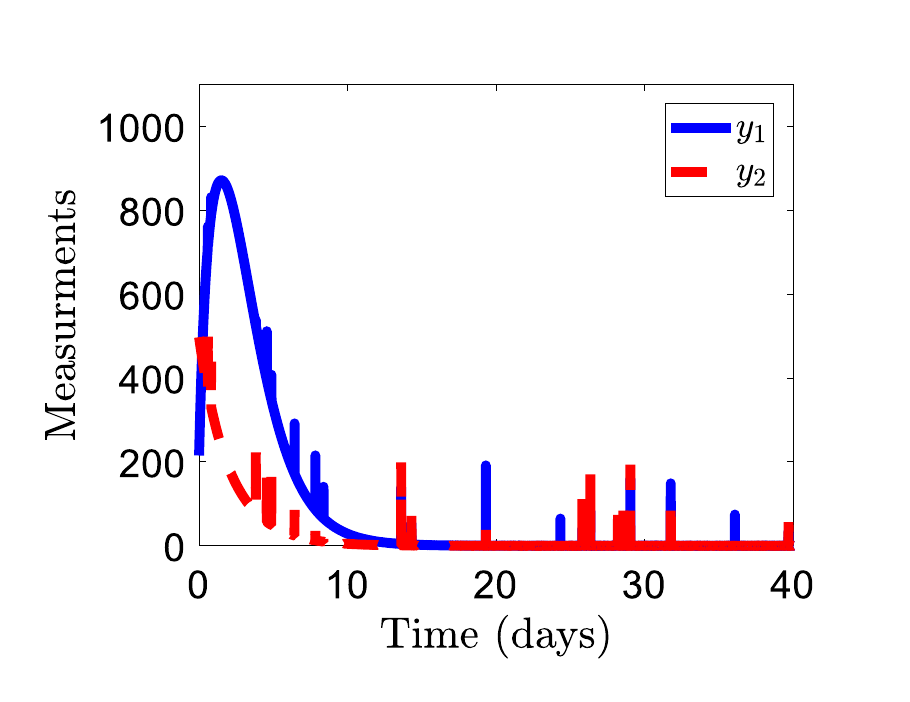}
	\end{subfigure}%
	\caption{State measurements ($z_2$,$z_3$) affected by a shot noise} \label{fig.In-meas}
	\vspace{-1em}
\end{figure}

\begin{figure*}[t]
	\centering \includegraphics[scale=0.25]{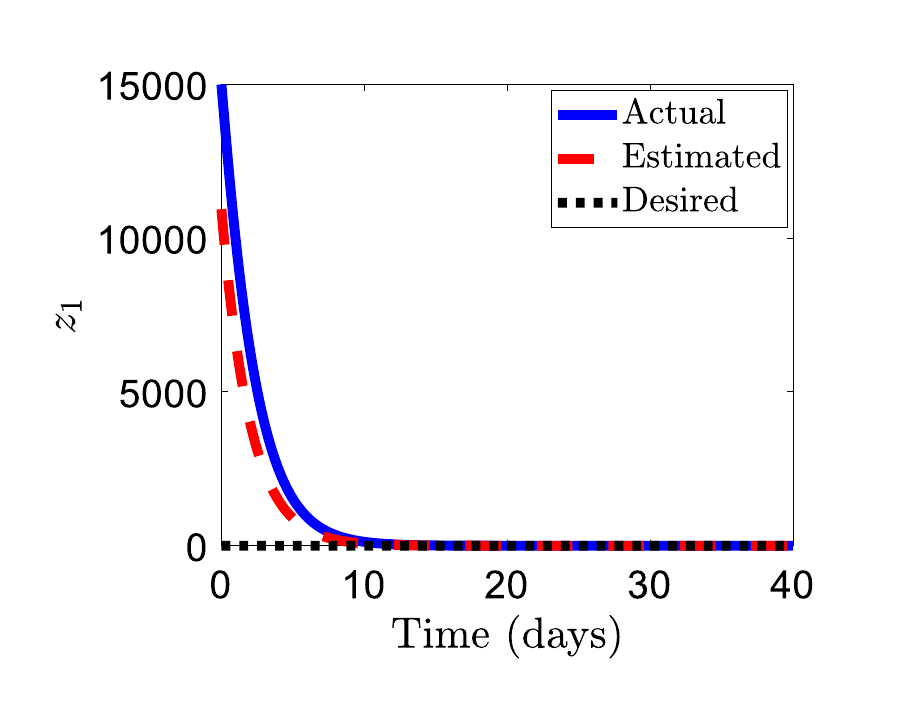}\hspace*{-0em}
	\includegraphics[scale=0.25]{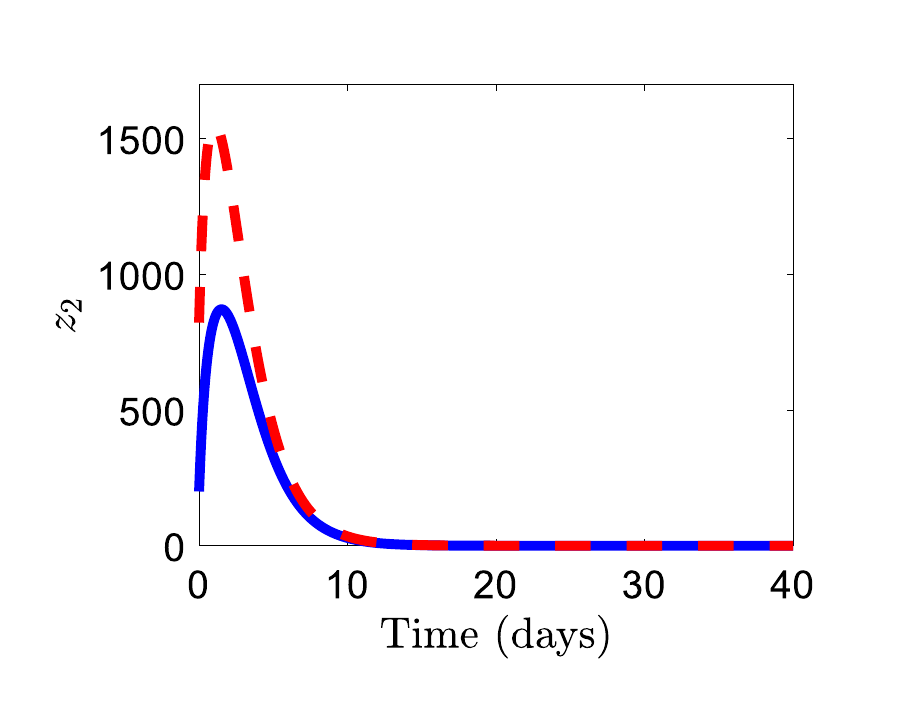}
	\includegraphics[scale=0.25]{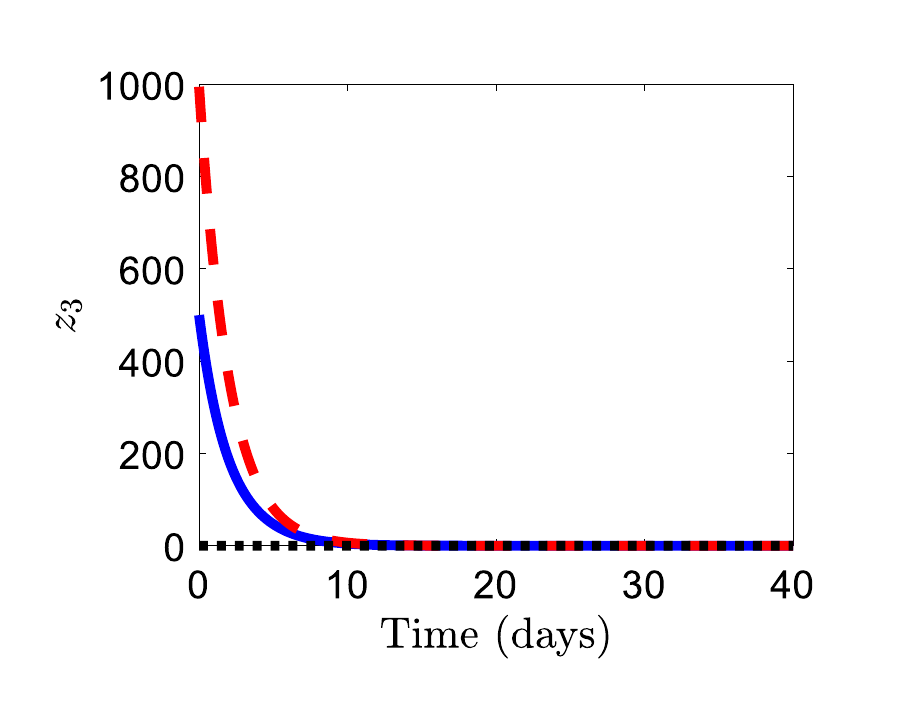}\hspace*{-0em}\\
	%\hspace*{12em}
	\includegraphics[scale=0.25]{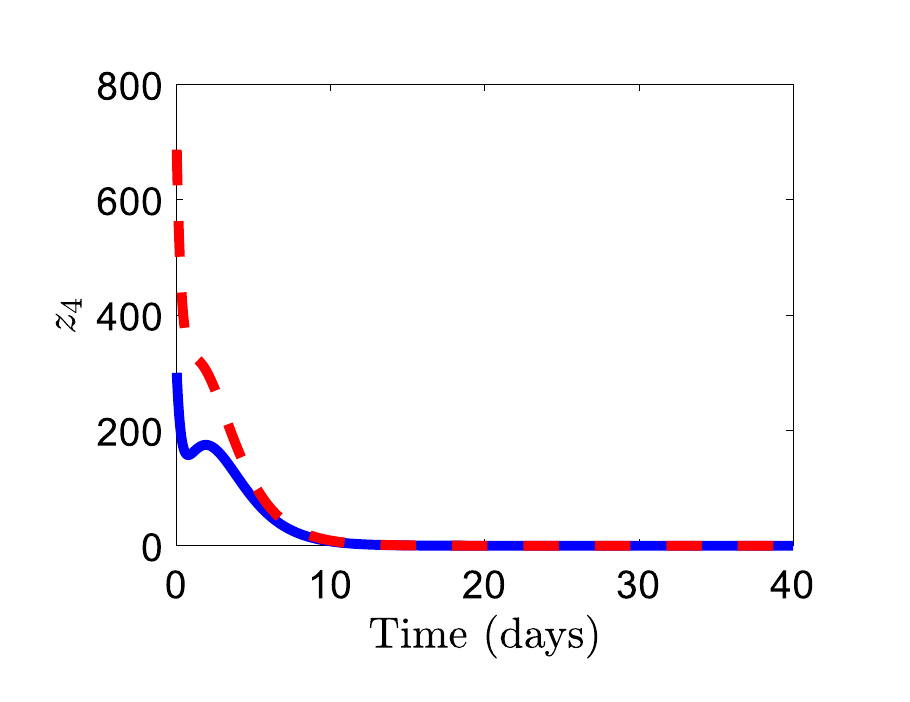}
	\includegraphics[scale=0.25]{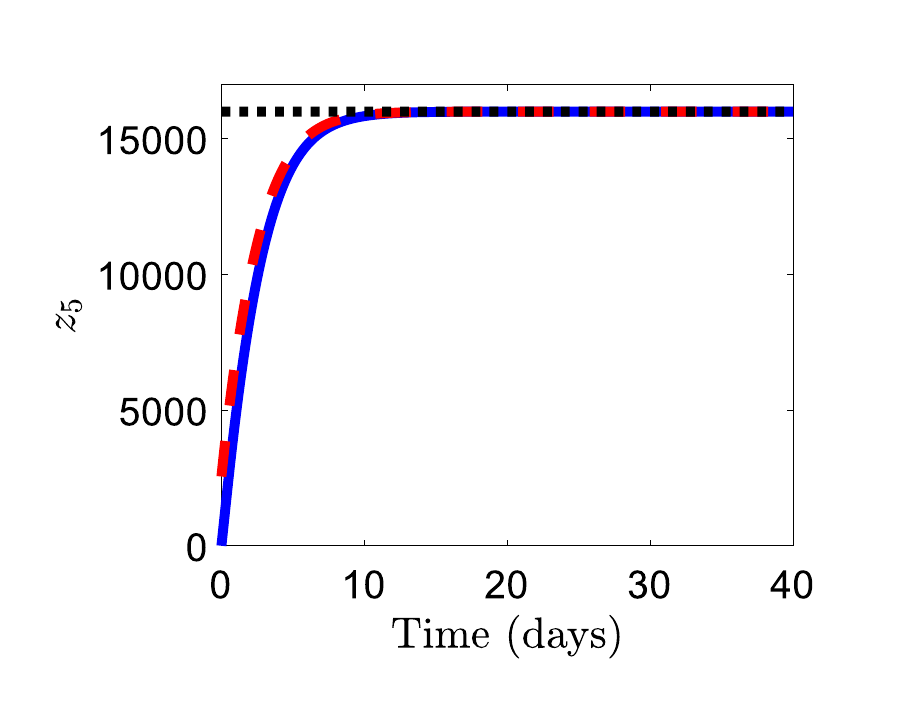}
	\caption{State estimation and tracking performance}\label{fig.trackest}
	\vspace{-1.5em}
\end{figure*}

\begin{figure}[t]
	\centering \includegraphics[scale=0.28]{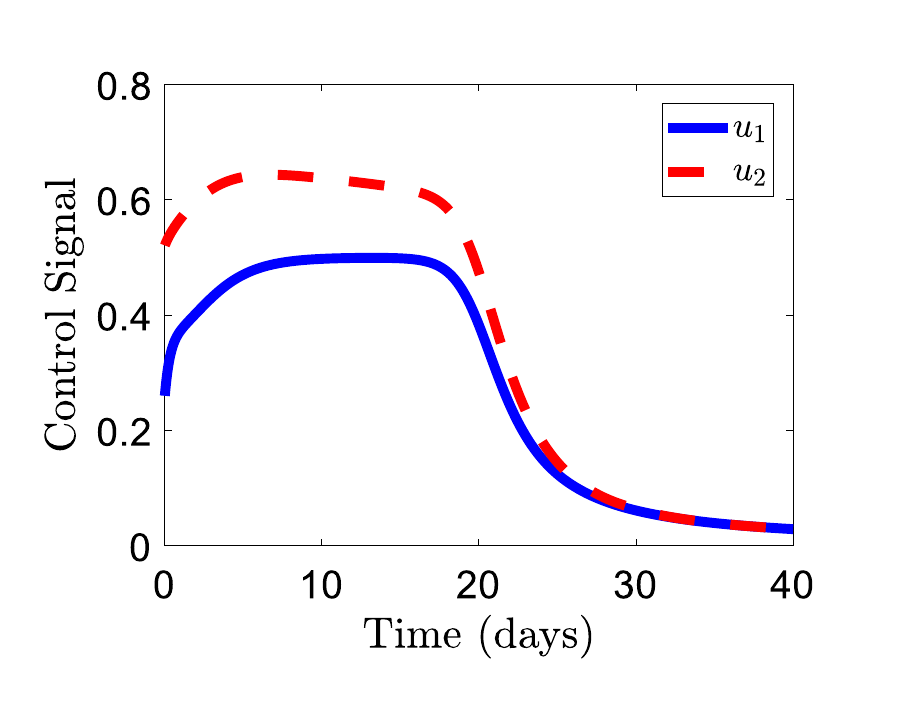}\hspace*{-0em}
	\includegraphics[scale=0.28]{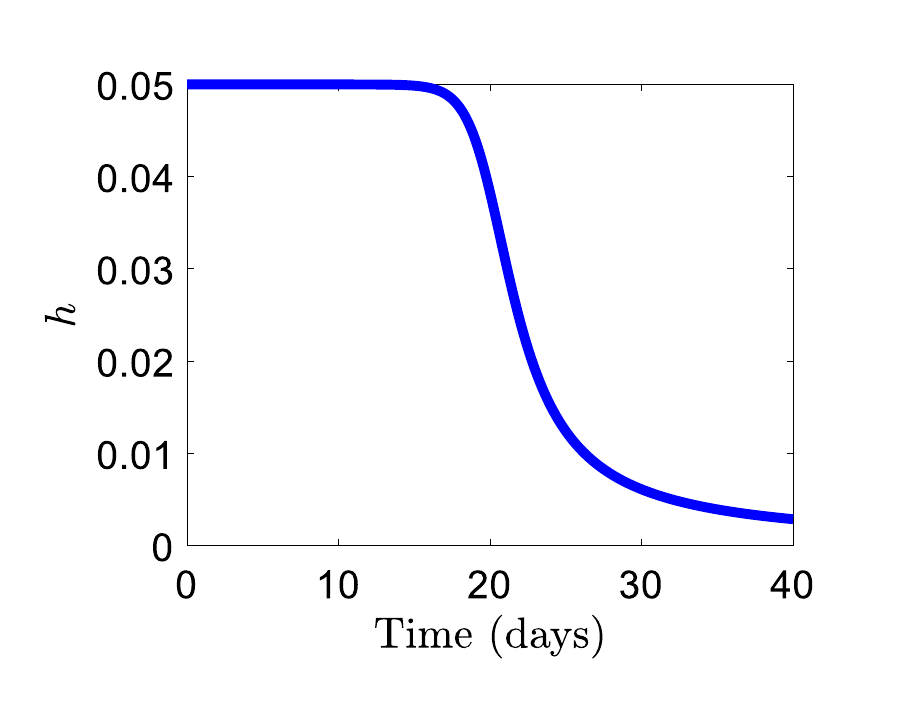}\hspace*{-0em}
	\caption{Control signals and RCLF constraint performance}\label{fig.cont}
	\vspace{-0em}
\end{figure}

The quadratic program-based CLF (QP-CLF) technique is a contemporary control approach that guarantees stability of closed-loop systems
while minimizing and bounding the control inputs~\cite{Ames2014,Ames2017}. However, modeling uncertainties and state estimation errors, i.e., $\Delta\neq0$, degrade the performance of such controllers~\cite{vahid1,vahid2}. To mitigate this issue, in this section, we aim to design a robust optimal controller by the unification of the EMCKF algorithm (\hyperref[section.est]{Section~\ref*{section.est}}) and the RCLF (\hyperref[section.contmin]{Section~\ref*{section.contmin}}) while utilizing the estimate of the system states. 
For this purpose, a QP optimization problem is employed to generate the same PWMC signal $u(\phi_{0_{rob}},\phi_{1})$, which enables the incorporation of
the RCLF constraint~\eqref{eq.RCLF} as well as the required control bounds while using the estimates of the states and the system parameters.

We begin by recovering the virtual input $\mu$ from the main control signal~\eqref{eq.lincont2} as 
\begin{align}\label{eq.lincont2Mu}
\mu=\hat{Y}-\dot{z}^d_e-\hat{Z}_{e}u.
\end{align}

To formulate the QP-RCLF-EMCKF controller while minimizing the virtual input $\mu$, the following cost
function should be minimized:
\begin{align}\nonumber
\label{eq.cost}
\mu^T\mu=&\hat{z}_{1}^2u_{1}^2+\hat{z}_{3}^2u_{2}^2
+2\hat{z}_{1}(\dot{z}^d_{e_1}-\hat{Y}_{1})u_{1}
+2\hat{z}_{3}(\dot{z}^d_{e_2}-\hat{Y}_{2})u_{2} \\ 
&-2(\hat{Y}_{1}\dot{z}^d_{e_1}+\hat{Y}_{2}\dot{z}^d_{e_2})
+\dot{z}^{d^2}_{e_1}+\dot{z}^{d^2}_{e_2}+\hat{Y}^2_{1}+\hat{Y}^2_{2}.
\end{align}

The control input $u$ has to be also restricted between
its prescribed minimum and maximum values such that
$\underline{u}\leq u_{i}\leq \bar{u}$, for $i=1,2$
with $\underline{u}=0$ and $\bar{u}=1$. Therefore,
a QP optimization problem with the aforementioned tracking and control
objectives can be formulated as:
\begin{align} \label{eq.QP0}\nonumber
x^*= &
\underset{x=(h,u)^T\in\Re^{3}}{\textrm{argmin}} \quad \mu^T \mu+ c h^2 \\
\nonumber
&\textrm{s.t.} \\
\nonumber &
\mathrm{RCLF \ constraint:} \ \ \phi_{1}^Tu+\phi_{0_{rob}} \leq h\\
 &
\mathrm{Control \ bound:} \ \ \ \ \ \underline{u}\leq u \leq \bar{u}
\end{align}
where $c$ is a relaxation coefficient for the RCLF
constraint~\eqref{eq.RCLF} when the control bound is enforced. Formally defining a
QP problem, the above optimization can be presented in the following form
\begin{align} \label{eq.QP}\nonumber
\mathbf{u^*} = &
\underset{\mathbf{x}\in\Re^{3}}{\textrm{argmin}} \quad
\frac{1}{2}\mathbf{u}^TH\mathbf{u}+B^T\mathbf{u}\\
\nonumber
&\textrm{s.t.} \\
\nonumber &
A_{1} \mathbf{u}\leq b_{1} \\
 &
A_{2} \mathbf{u}\leq b_{2}
\end{align}
with
\begin{align} \label{eq.HB}
H= 2\left[ \begin{array}{ccc}
c &0 & 0 \\0 & \hat{z}_{1}^2 & 0\\ 0 & 0 & \hat{z}_{3}^2
\end{array} \right], \quad
B = 2\left[ \begin{array}{c}
0 \\ \hat{z}_{1}(\dot{z}^d_{e_1}-\hat{Y}_{1}) \\ \hat{z}_{3}(\dot{z}^d_{e_2}-\hat{Y}_{2})
\end{array} \right]
\end{align}
and
\begin{align} \label{eq.AB}\nonumber
A_{1}&= \left[ \begin{array}{cc}
-1 & \phi_{1}^T
\end{array} \right], \quad
b_{1} = -\phi_{0_{rob}} \\ 
A_{2}&= \left[ \begin{array}{ccc}
0 & 1 & 0\\ 0 & 0 & 1 \\ 0 & -1 & 0\\ 0 & 0 & -1
\end{array} \right], \quad
b_{2} = \left[ \begin{array}{c}
\bar{u}_{1}\\  \bar{u}_{2} \\ \underline{u}_{1}\\  \underline{u}_{2}
\end{array} \right].
\end{align}

\begin{figure}[!htb]
	\centering \includegraphics[scale=0.25]{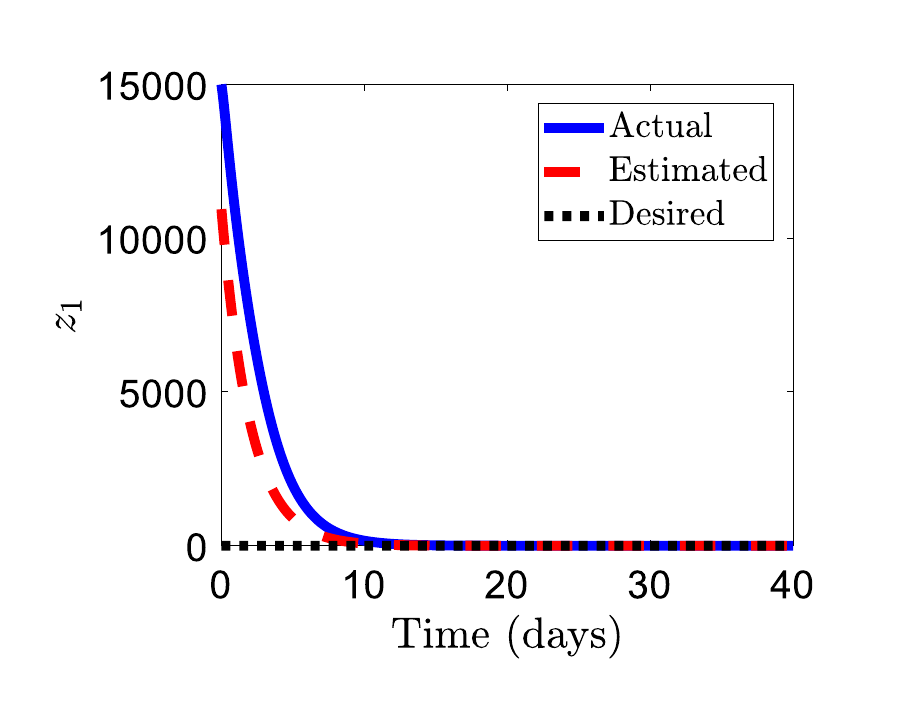}
	\includegraphics[scale=0.25]{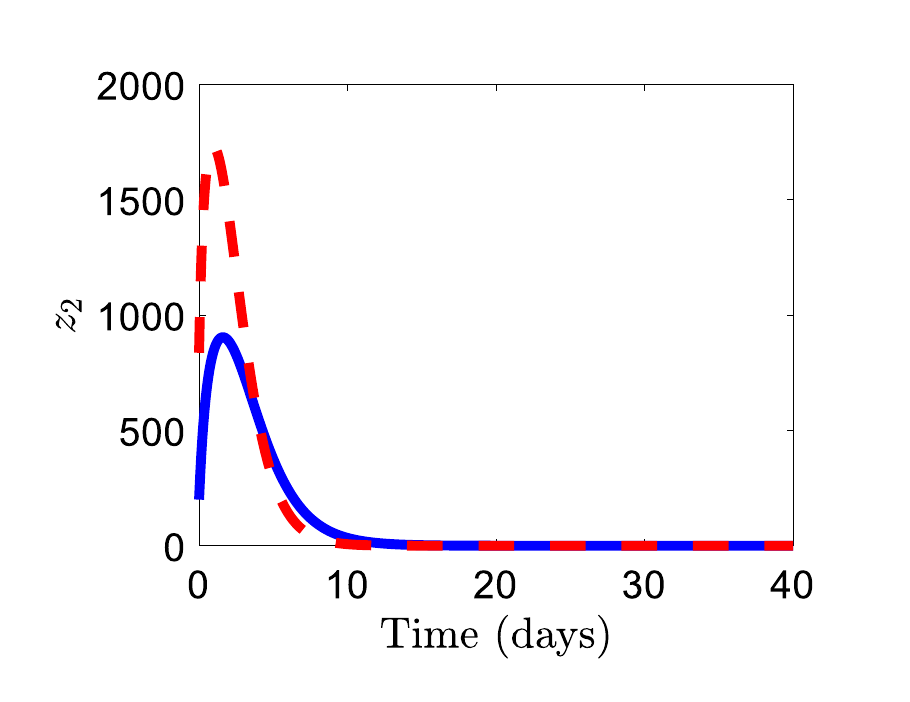}
	\includegraphics[scale=0.25]{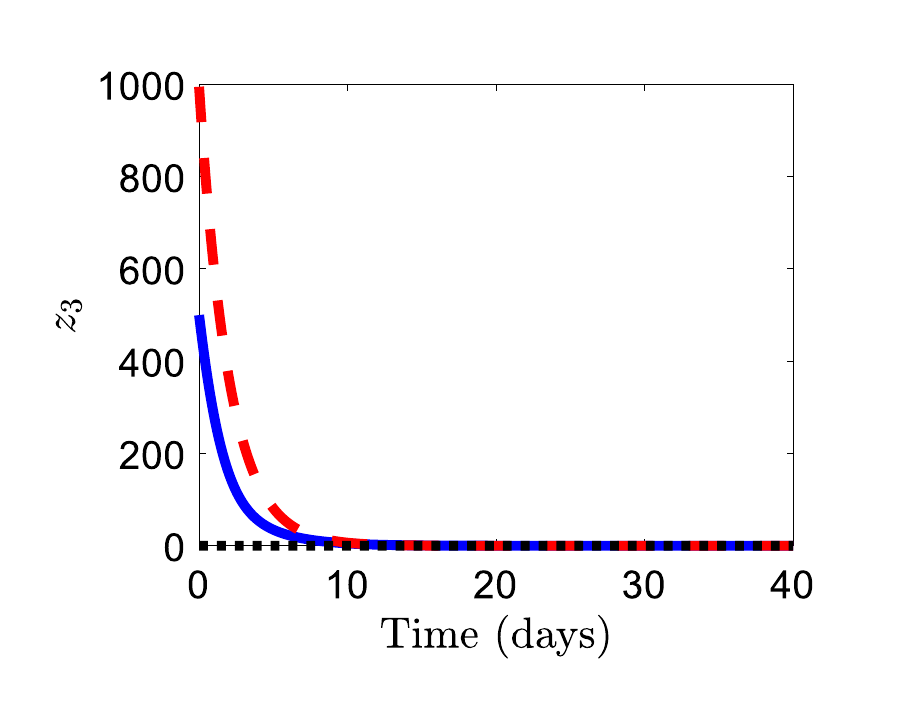}\\
	\includegraphics[scale=0.25]{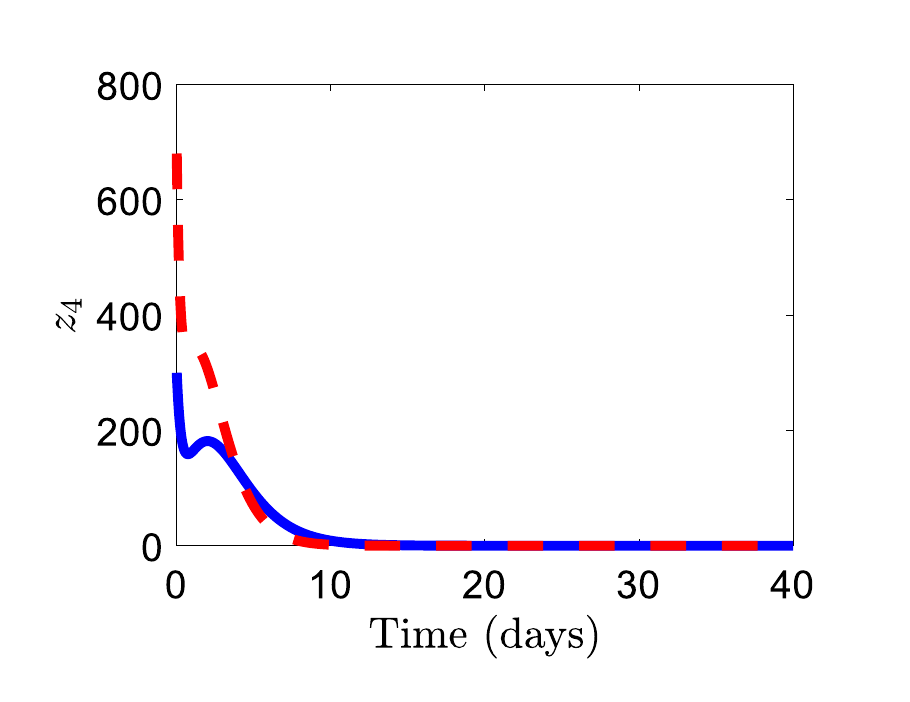}
	\includegraphics[scale=0.25]{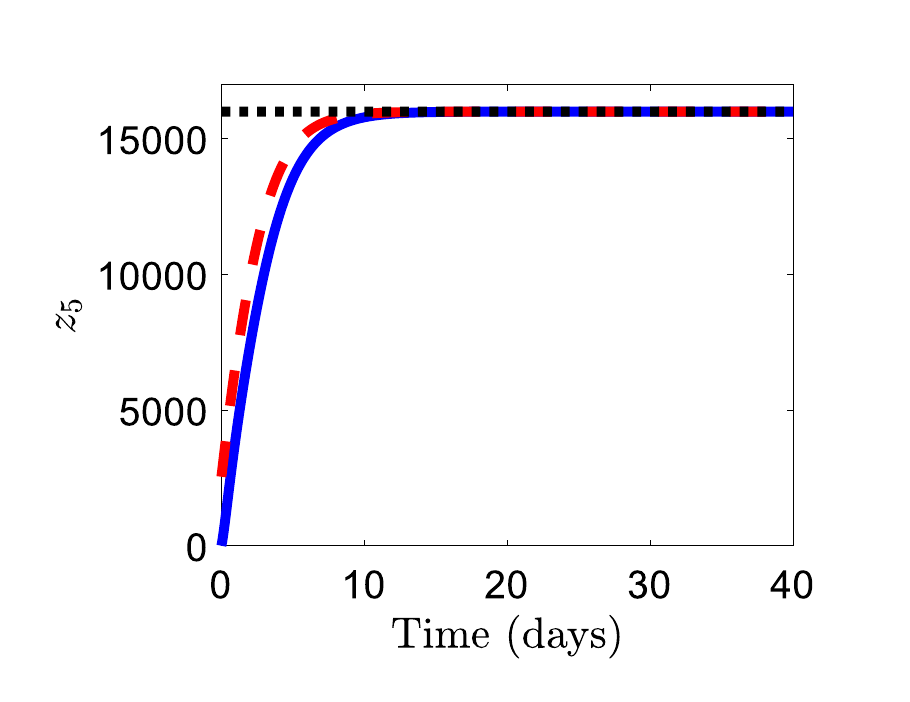}\hspace*{-0em}
	\\
	(a) $\Delta\Theta=+50\%$
	\\
	 \includegraphics[scale=0.25]{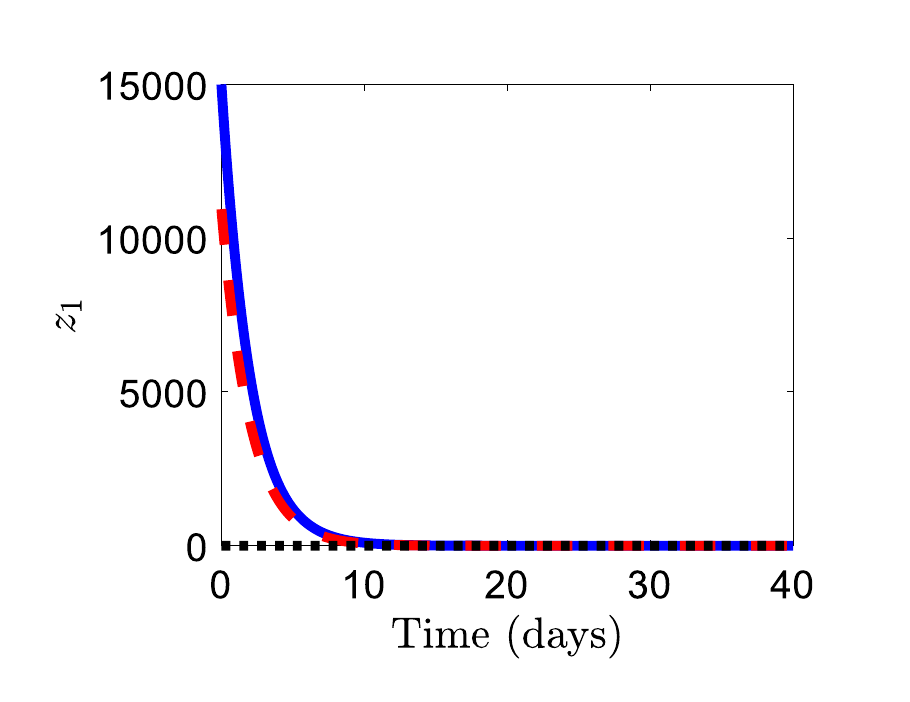}\hspace*{-0em}
	\includegraphics[scale=0.25]{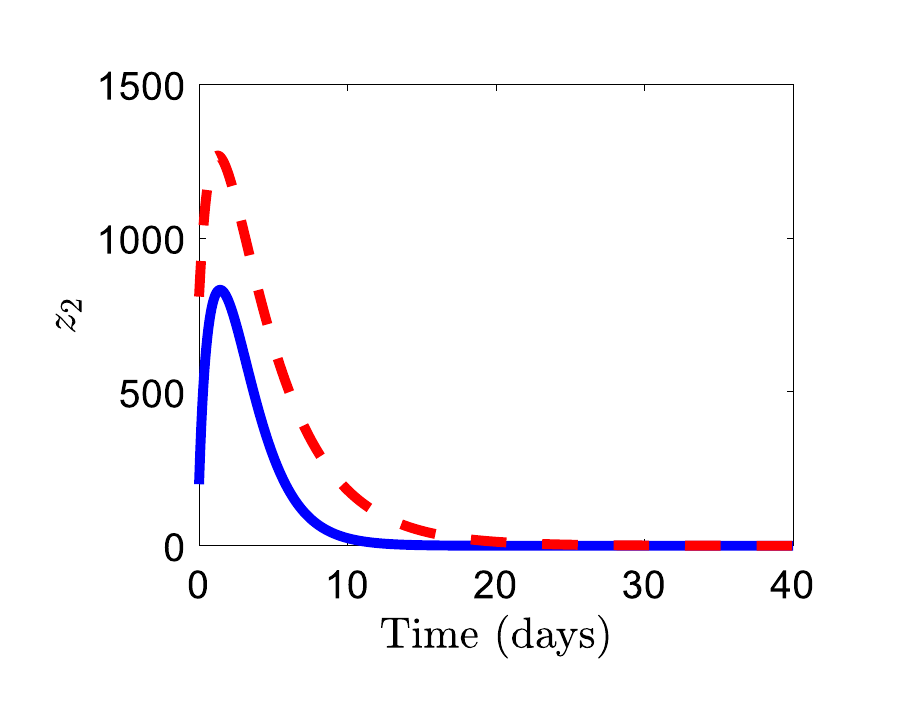}\hspace*{-0em}
	\includegraphics[scale=0.25]{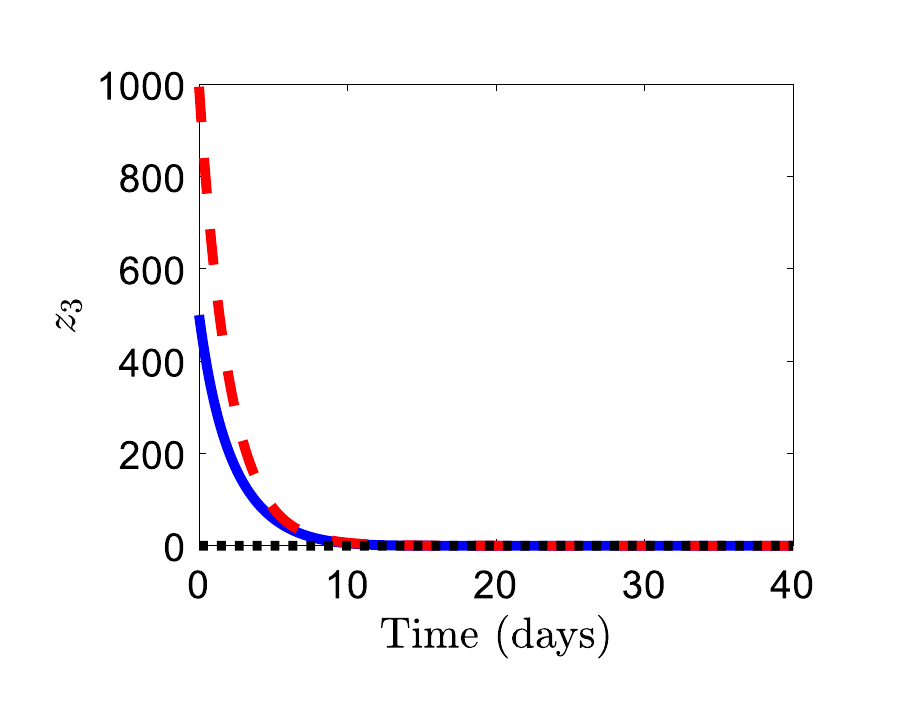}\hspace*{-0em}
	\\
	\includegraphics[scale=0.25]{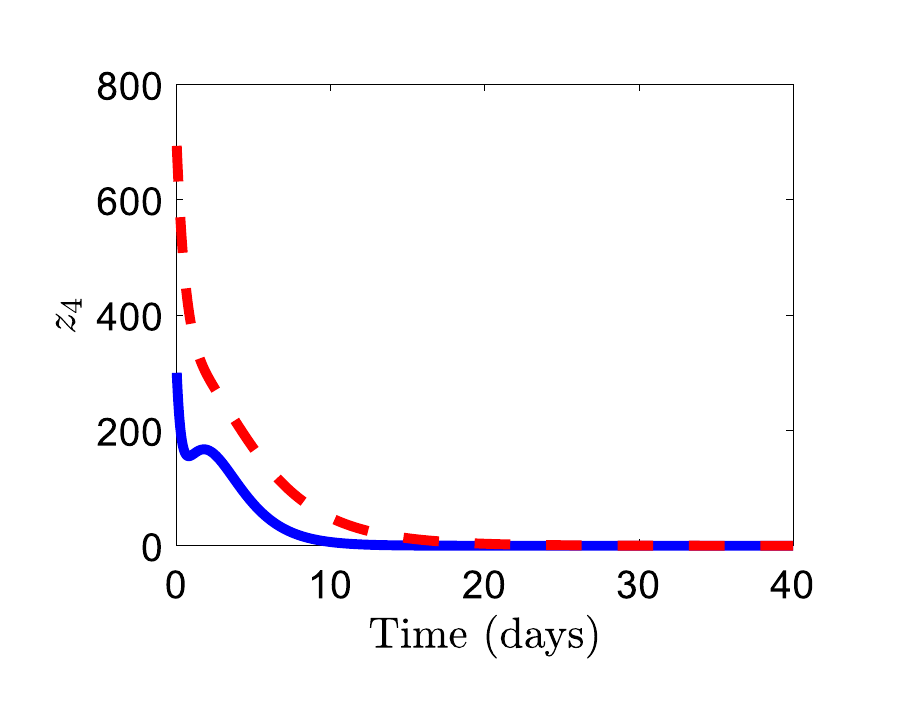}\hspace*{-0em}
	\includegraphics[scale=0.25]{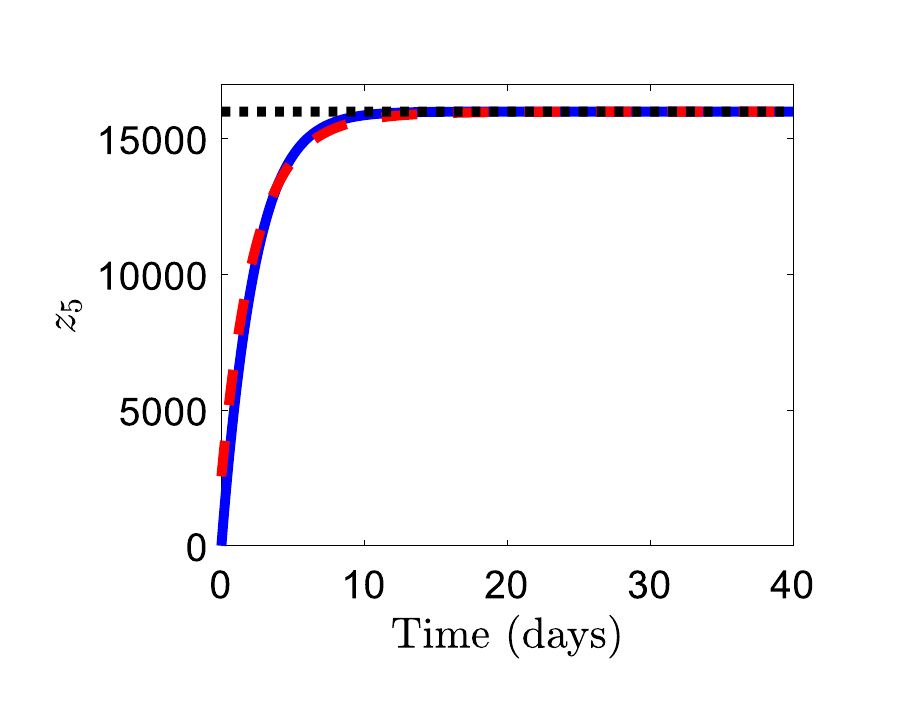}\hspace*{-0em}
	\\
	(b) $\Delta\Theta=-50\%$
	\vspace*{1em}
	\caption{State estimation and tracking performance under $\pm50\%$ parameter uncertainty}\label{fig.trackest50}
	\vspace{-0.5em}
\end{figure}

\begin{figure}[!htb]
	\hspace*{4em} \includegraphics[scale=0.28]{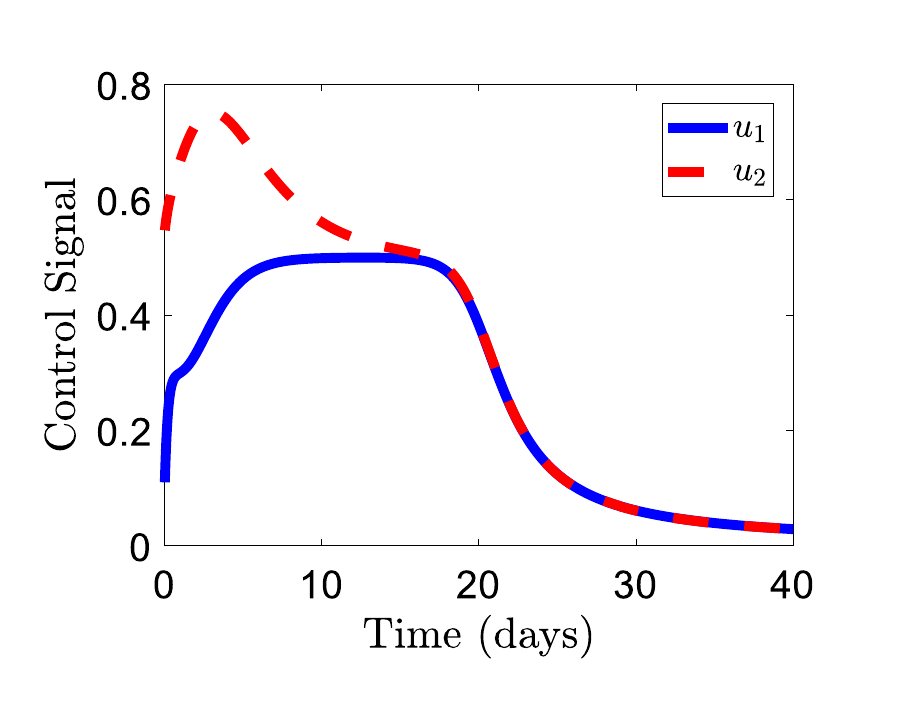}\hspace*{-0em}
	\includegraphics[scale=0.28]{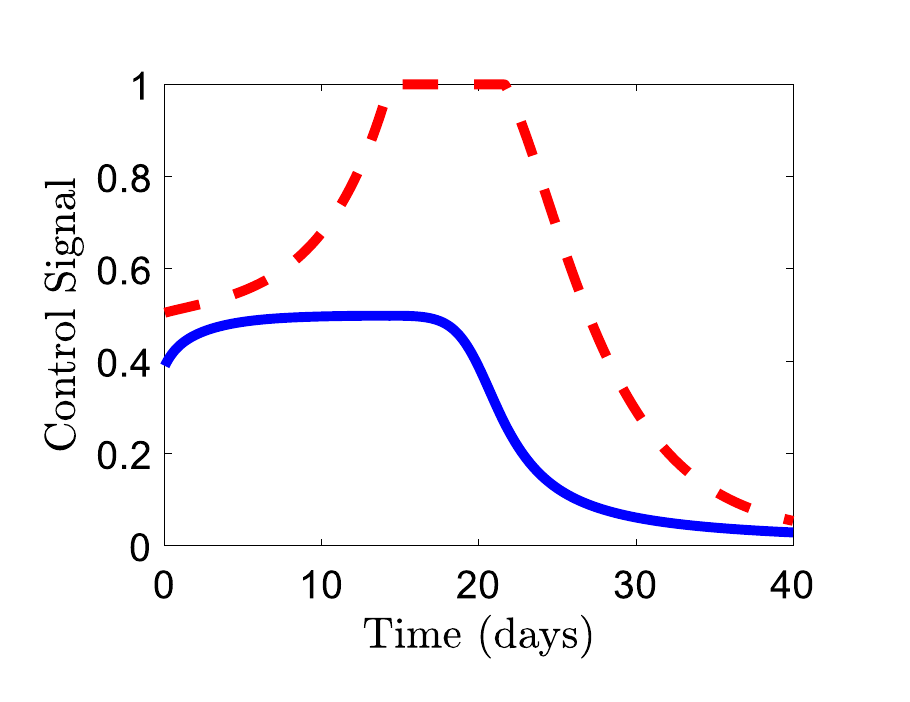}\hspace*{-0em}
	\\
	\begin{tabular}{cc}
		\hspace*{6em} (a) $\Delta\Theta=+50\%$ &
		\hspace*{4em} (b) $\Delta\Theta=-50\%$
	\end{tabular}
	\vspace*{1em}
	\caption{Vaccination rate $u_1$ and antiviral treatment rate $u_2$ under $\pm50\%$ parameter uncertainty}\label{fig.cont50}
	\vspace{-0em}
\end{figure}

Equations~\eqref{eq.QP}, ~\eqref{eq.HB},
and ~\eqref{eq.AB} show that the proposed controller uses the estimated
states (by using EMCKF algorithm in \hyperref[section.est]{Section~\ref*{section.est}}),
the estimated parameters (best guess), tracking error, and the first derivative
of the desired trajectory as a four-tuple $(\hat{z},\hat{\theta},e,\dot{z}^d_e)$.
The general structure of the proposed QP-RCLF-EMCKF for
the influenza epidemics in an interactive human society is illustrated in
Fig.~\ref{fig.FluFlo}.

\section{Simulation Results}\label{section.Sim}

In this section, the proposed control methodology QP-RCLF-EMCKF is implemented on the influenza
epidemic model~\eqref{eq.model} whose parameters
are shown in Table~\ref{table.syspara2}. We aim to minimize the susceptible and infected individuals in an interactive human society with population of 16000. The initial value of the
state variables is considered 
\begin{equation*}
z(0)=[15000,200,500,300,0]^T
\end{equation*} 
that is assumed to be different from the initial value of the filter states
\begin{equation*}
\hat{z}(0)=[11000,800,1000,700,2500]^T.
\end{equation*}

Note that the summation of initial state variables is equal to the population
of the main society.
The simulation runs for 40 days. Table~\ref{table.syspara} provides the design parameters
of the proposed approach for the state estimation algorithm explained in
~\hyperref[section.est]{Section~\ref*{section.est}} and the controller formulated in
~\hyperref[section.cont]{Section~\ref*{section.cont}}. The design parameters are
tuned to provide a good performance of the proposed approach.

The effects from the other human societies on the main interactive society is
modeled by a shot noise. Thus, the measurement noise is regarded as a non-Gaussian noise,
which is a Gaussian noise that is affected by a shot noise as described in Eq.~\eqref{eq.noise}.
In the simulation, the shot noise is seen as 20 impulses with magnitude of 200,
which is randomly enforced to the measurement noise. This shot noise models the random
entrance of 200 exposed and infected individuals from the other human societies
into the human society of population 16000.
Thus, the measurements ($z_2$,$z_3$) are affected by these 200 individuals during the
simulation as shown in Fig.~\ref{fig.In-meas}.

\subsection{State estimation, tracking performance, and control effort} \label{section.simest}

Figure~\ref{fig.trackest} shows the state
estimation performance for the influenza epidemics along with the convergence of populations ${z}_{1}$ and ${z}_{3}$. It is seen that the proposed EMCKF
algorithm is able to accurately estimate the state variables while only measuring the
populations ${z}_{2}$ and ${z}_{3}$. This accurate estimation is
achieved when the shot noise is enforced to the measurement noise, which
represents an impulsive random entrance of the exposed and infected populations to the main human
society of 16000.
This implies that the proposed estimation algorithm has a strong robustness when the system
is perturbed by non-Gaussian noises.

Figure~\ref{fig.trackest} also shows that the susceptible ${z}_{1}$ and infected ${z}_{3}$
individuals of the interactive human society are minimized in 14 days under the proposed control strategy. The convergence
of variables ${z}_{1}$ and ${z}_{3}$ results in the convergence of populations ${z}_{2}$
and ${z}_{4}$, and in turn the entire population ${z}_{5}$ is recovered.
This implies that
the  proposed controller is able to recover all individuals of the human society
with the population of 16000, even when the external infected individuals from other
societies randomly invade the main society during a treatment time of 40 days. These results are in agreement with our main results presented in \hyperref[section.contmin]{Section~\ref*{section.contmin}} and \hyperref[th.Main2]{Theorem~\ref*{th.Main2}} based on which UUB/convergence of system's errors is
guaranteed.

Figure~\ref{fig.cont} illustrates the rate of vaccination for
susceptible individuals ${u}_{1}$ and the rate of antiviral treatment for the infected
individuals ${u}_{2}$.
It is seen that the control signals generated by the proposed control technique
fairly decreases to zero at the end of the treatment time.
It can be also noted that none of the control signals hit
the maximum control bound $\bar{u}$ as the peak controls are
$u_{1_{max}}=0.49$ and $u_{2_{max}}=0.64$.
Figure~\ref{fig.cont} also demonstrates the RCLF constraint violation
during the simulation. It is seen that the RCLF violation is bounded by 0.05
when the relaxation coefficient is tuned as $c=10$. A smaller value of $c$ relaxes
the RCLF constraint and decreases the possibility of its conflict with the control bound
constraint; however, smaller $c$ increases $h$ and in turn deteriorates the tracking
performance.
For higher relaxation coefficient $c$, $h$ is relatively zero and the RCLF constraint is never
violated, but the QP may be infeasible due to the conflict of the RCLF constraint with the
control bounds. Thus, the penalty coefficient $c$ should be carefully selected to make a trade off
between the tracking performance and the control constraints.

\subsection{Robustness to parameter uncertainty} \label{section.robust}

Different societies and populations can
result in the influenza model~\eqref{eq.model}
with different values of the system parameters $\Theta$.
To evaluate the robustness of the proposed control scheme against the parameter
perturbation, the system parameters are deviated by $\pm50\%$ from their
nominal values. Figure~\ref{fig.trackest50} illustrates the state estimation
and tracking performance of the influenza epidemics when the system parameters are perturbed
by $\pm50\%$. It is seen that the proposed EMCKF algorithm can still provide an accurate
state estimation under either case. Under $+50\%$ parameter perturbation, the number of susceptible $z_1$ and infected $z_3$ populations converges to a small ultimate ball around zero in 14 days using the proposed controller. In case that
$\Delta\Theta=-50\%$, although the estimated states $z_2$ and $z_4$ have a
sluggish convergence to the actual states, the EMCKF algorithm can render a
general convenient estimation performance. In this case, the convergence of
$z_1$ and $z_3$ is also achieved in the same days as of $\Delta\Theta=0\%$
and $\Delta\Theta=+50\%$. This demonstrates that the proposed approach achieves good
robustness against the parameter perturbation. These findings support the claim of our main results presented in \hyperref[th.Main2]{Theorem~\ref*{th.Main2}} in which UUB/convergence of the tracking errors is ensured even in the presence of parameter uncertainties and state estimation error. 

Figure~\ref{fig.cont50} shows the control signals under $\pm50\%$
parameter uncertainty. It is observed that the rate of vaccination
for $z_1$ ($u_1$) under both cases $\Delta\Theta=+50\%$ and $\Delta\Theta=-50\%$
has relatively similar magnitude and behavior compared to $u_1$ in no
perturbation case. Under both $\Delta\Theta=+50\%$ and $\Delta\Theta=-50\%$,
the maximum value of $u_1$ is $u_{1_{max}}=0.49$. However, the rate of antiviral treatment
for $z_3$ ($u_2$) under $\Delta\Theta=+50\%$ meets a higher magnitude in the
first 10 days ($u_{2_{max}}=0.75$), which is $17\%$ higher than $u_{2_{max}}$
in the case of no perturbation. Under $\Delta\Theta=-50\%$, although $u_2$ hits the
control bound $\bar{u}=1$ during $t\in[15, 22]$, convergence of $z_3$ is maintained.
This implies that there is no conflict between the control bounds and the RCLF
constraint such that they can be achieved at the same time. This demonstrates that the proposed
approach is able to achieve convergence of system solutions and to satisfy
the constraints in the presence of parameter perturbation
and state estimation errors.

\subsection{Superiority of the EMCKF algorithm over the ordinary EKF for the influenza epidemics} \label{section.ComEKF}
\begin{figure*}[!htb]
\centering \includegraphics[scale=0.25]{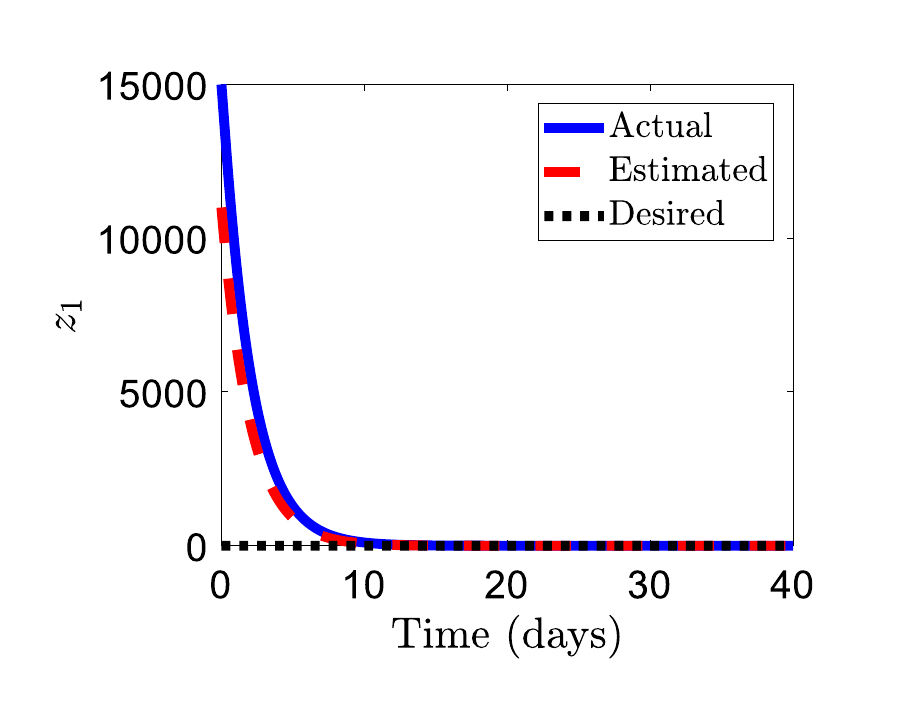}\hspace*{-0em}
\includegraphics[scale=0.25]{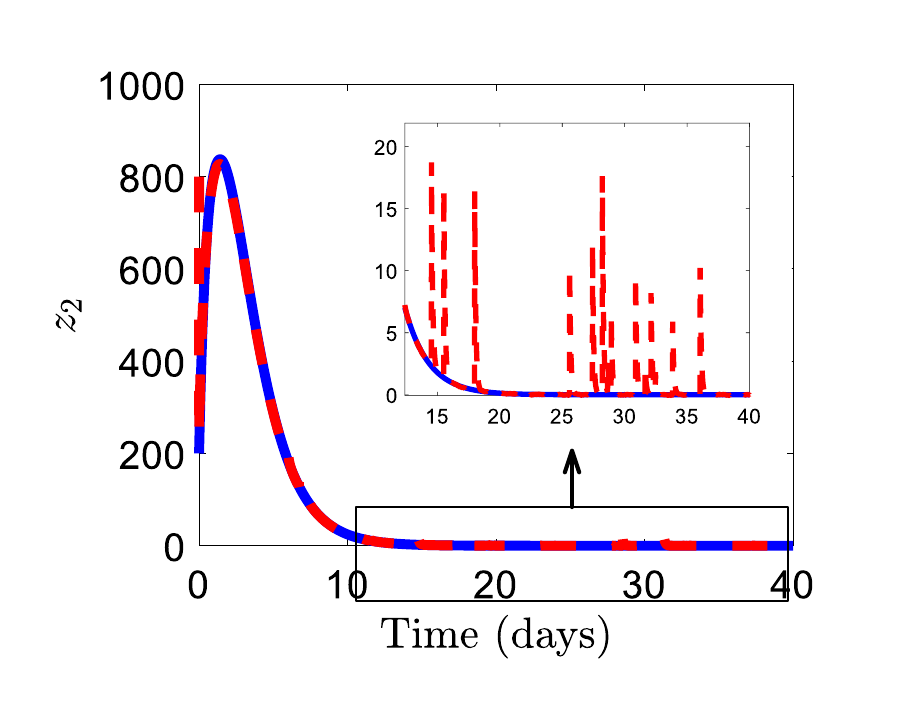}\hspace*{-0em}
\includegraphics[scale=0.25]{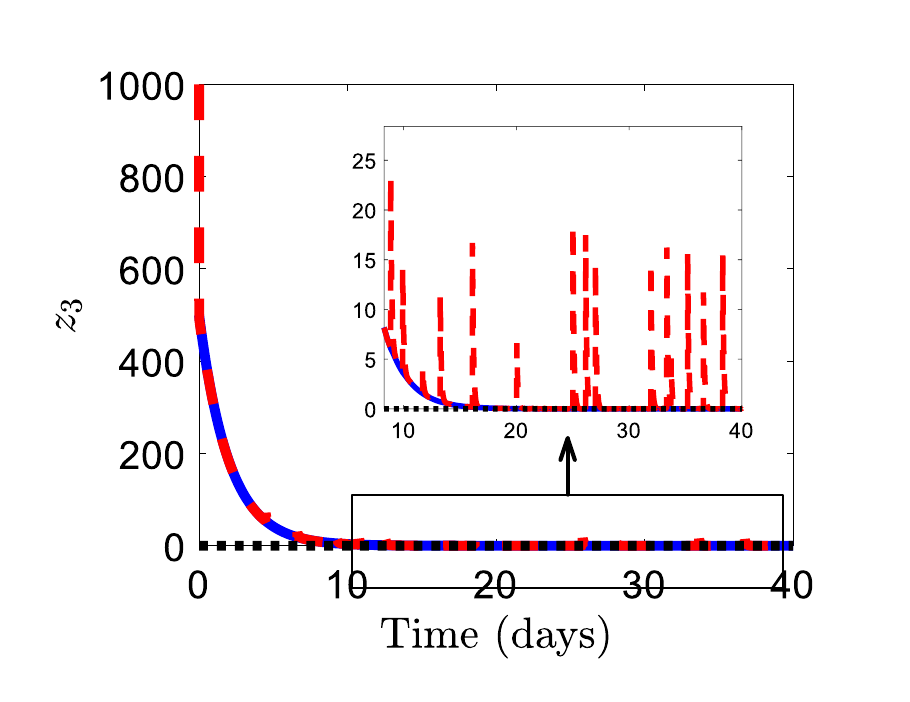}\hspace*{-0em}
\\

\includegraphics[scale=0.25]{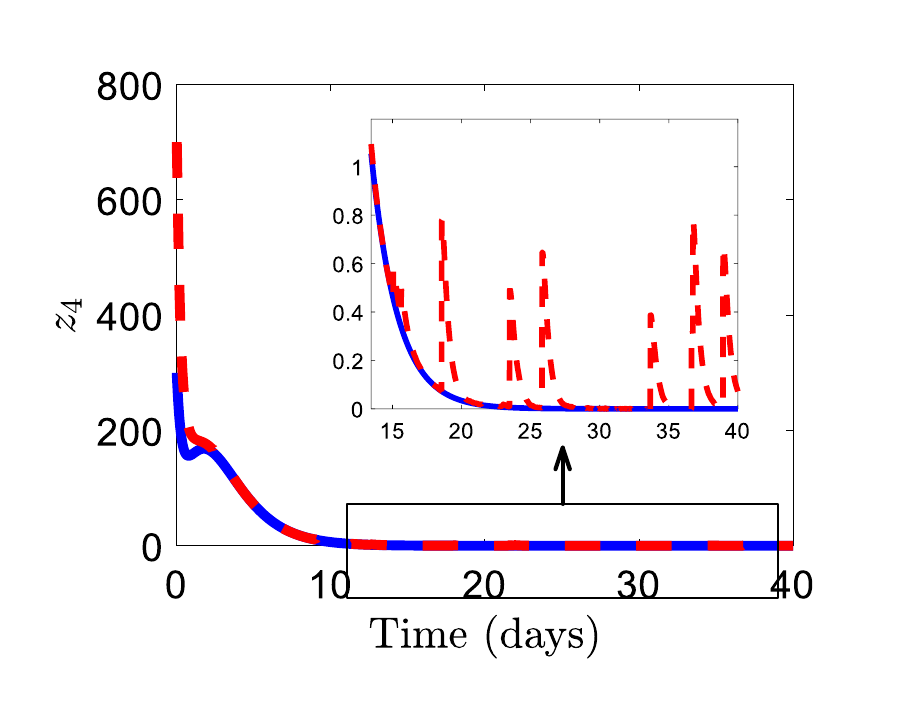}\hspace*{-0em}
\includegraphics[scale=0.25]{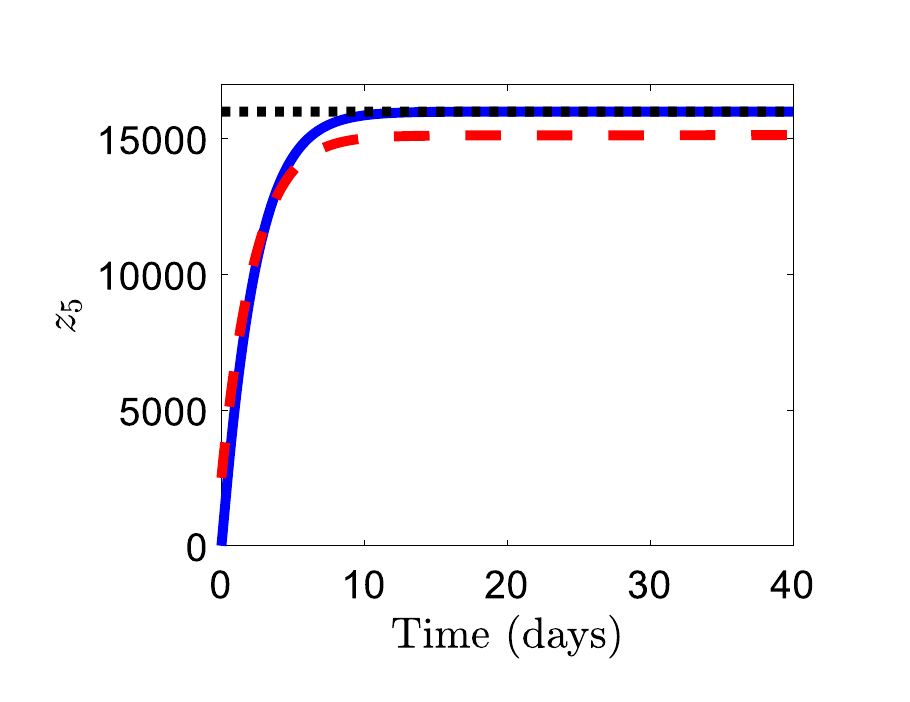}\hspace*{-0em}
\caption{State estimation and tracking performance using the ordinary EKF}\label{fig.trackestEKF}
\vspace{-0.5em}
\end{figure*}
\begin{figure}[!htb]
\centering \includegraphics[scale=0.28]{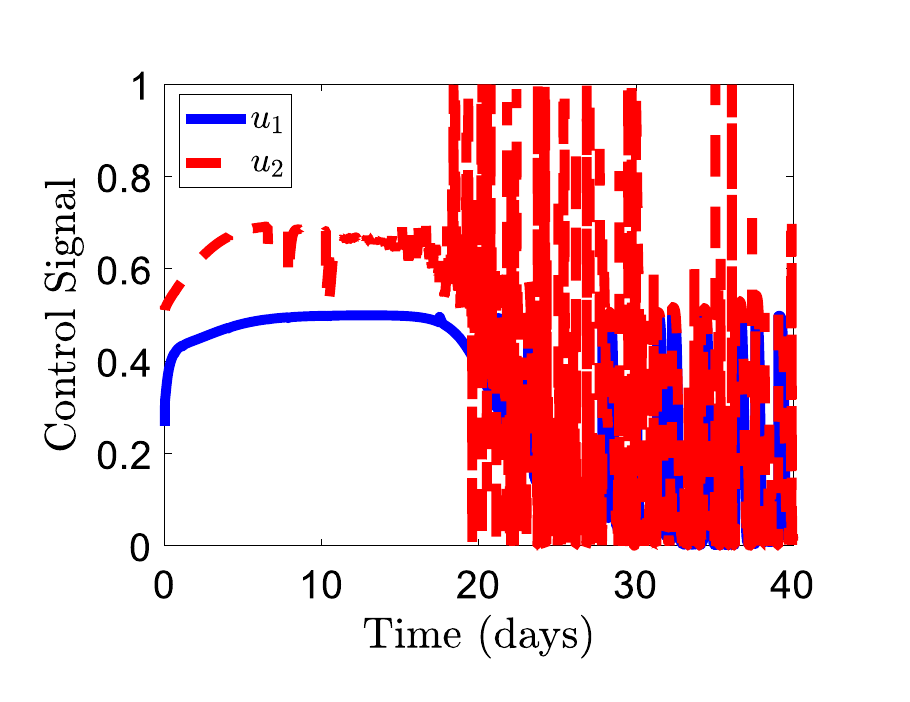}\hspace*{-0em}
\includegraphics[scale=0.28]{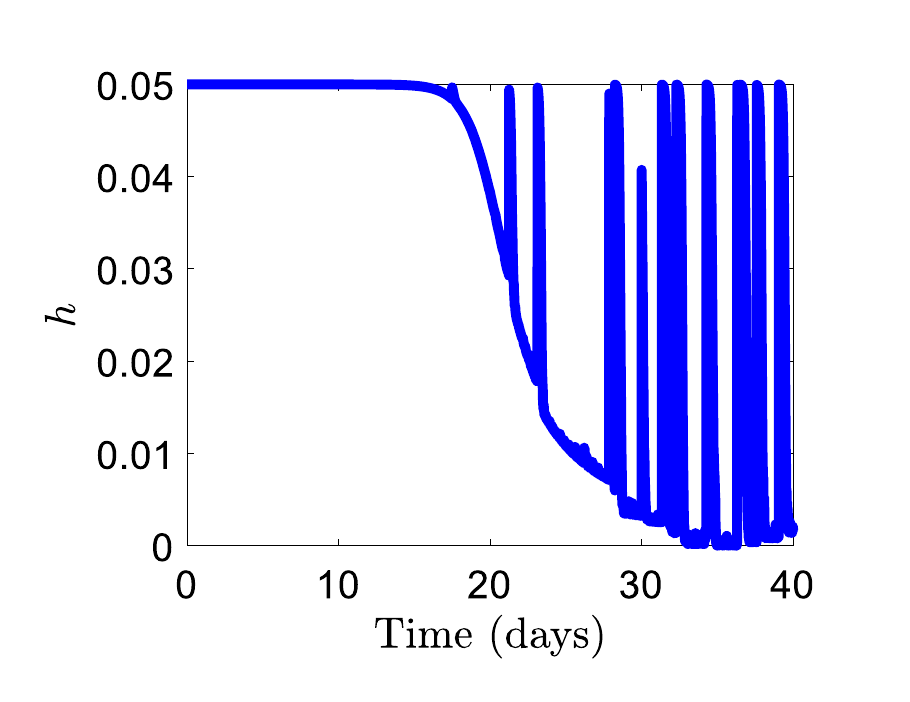}\hspace*{-0em}
\caption{Control signal and RCLF constraint performance using the ordinary EKF}\label{fig.contEKF}
\vspace{-0em}
\end{figure}
In this section, we highlight superiority of the EMCKF algorithm over
the ordinary EKF when the system is affected by the shot noise introducing the random entrance of exposed and infected individuals
from other societies to the society of interest.
Figure~\ref{fig.trackestEKF} illustrates the state estimation and tracking performance
of the influenza epidemic system under the proposed controller but when an ordinary
EKF is employed. It is seen that the estimated states $z_2$, $z_3$, and $z_4$ contain unpleasant impulses
stemming from the shot noise and in turn do not converge to their actual states.
This results in a steady state estimation error for the state $\hat{z}_5$.
Thus, it is seen that the estimation performance deteriorates
when the system is disturbed by the shot noise and the EKF is employed.

Since the proposed controller uses the estimated states,
inconvenient state estimation of the EKF negatively impacts the generated
control signals as shown in Fig.~\ref{fig.contEKF}. Both the rate of
vaccination for $z_1$ and the rate of antiviral treatment for $z_3$ intensively chatter after
day 20 and even $u_2$ hits
the control bound $\bar{u}$. This shows that improper estimation performance of the
EKF in the presence of shot noise causes
the control signal chattering, resulting in higher control cost.
Figure~\ref{fig.contEKF} also shows that the RCLF constraint violation is not smooth and
chatters after day 20. This demonstrates that the proposed controller can not preserve its robustness for an interactive human society (when the main human society is not isolated from the other societies i.e., existing of non-Gaussian noise) when the ordinary EKF is employed instead of the proposed EMCKF algorithm.

\begin{comment}
Table~\ref{table.numres} lists root mean square error ($\mathrm{RMSE}$) values for tracking $\textrm{RMSE}_t$ and estimation $\textrm{RMSE}_e$, RMS of the control signals $\mathrm{RMS_u}$, and
maximum control signal value $u_{\textrm{max}}$ using the proposed controller. As seen from the table, although parameter perturbation significantly deteriorates the state estimation performance, the tracking cost $\textrm{RMSE}_t$ increases by $33\%$ for $\Delta\Theta=+50\%$ and decreases by $28\%$ for $\Delta\Theta=-50\%$.
\begin{table}[t]
\centering
\scriptsize
\caption{Numerical results for state estimation $\mathrm{RMSE_{e}}$ (whole simulation, steady state for $t\in[25, 40]$), tracking performance $\mathrm{RMSE_{t}}$ (whole simulation, steady state), and control effort $\mathrm{u_{max}}$ ($\mathrm{u_{1_{max}}}$, $\mathrm{u_{2_{max}}}$).}
\label{table.numres}
\centering
\begin{tabular}{cccccc}
\hline
 & $\mathrm{RMSE_{t}}$ & $\mathrm{RMSE_{e}}$ & $\mathrm{RMS_{u}}$ & $\mathrm{u_{max}}$ \\
\hline
\hline
\centering
Nominal &             (2652, 0.21) & (1700, 0.17) & 0.79 & (0.49, 0.64) \\
\centering
$\Delta\Theta=+50\%$& (2874, 0.28) & (2244, 0.67) & 0.77 & (0.49, 0.75) \\
\centering
$\Delta\Theta=-50\%$& (2469, 0.15) & (1379, 3.29) & 1 & (0.49, 1)
\end{tabular}
\vspace{-1em}
\end{table}
\end{comment}
\section{Discussion, Conclusions, and Future Work}\label{section.Conc}
\subsection{Discussion} \label{section.Conc1}
Control of influenza
epidemics in a human society is an important global
health concern that imposes economic and epidemiological burdens.
The optimal control strategy is one of the most popular design approaches
that has been employed to control the influenza epidemics.
However, previous optimal control approaches have been designed with the assumptions
of fully-known dynamics and fully-measurable states in addition to considering an isolated human society.
The adaptive control strategy is an efficient design method for controlling the influenza
epidemics in the presence of dynamic uncertainties. To cope with the modeling inaccuracies, an adaptive control method has been recently designed in~\cite{Sharifi}
while still assuming that the system's states are measurable and the human society is isolated. In addition, that controller
did not take the optimality of the vaccination and antiviral
treatment rates into account. 

Since the influenza dynamic models are a set of
nonlinear differential equations, the EKF is a convenient algorithm for the state
estimation of such systems. However, since the human society of interest is not isolated
from the other societies (it is an interactive society that is impacted by non-Gaussian noise), performance of the ordinary EKF
deteriorates in the presence of other societies' interactions.

\subsection{Conclusions} \label{section.Conc2}
Motivated by the aforementioned shortcomings of the existing works applied for the influenza epidemics and the aim of devising a new multi-objective controller for such systems, this paper presented a state estimation-based robust optimal control strategy for the influenza epidemics in an interactive human society in the presence of modeling uncertainties. An EMCKF algorithm was presented for state estimation purpose and a QP optimization problem was formulated w.r.t. a RCLF to recover the entire population of an interactive human society while compensating the state estimation error and the modeling error in an optimal fashion.  
The proposed QP-RCLF-EMCKF controller achieved multiple design specifications such
as state estimation, tracking, control optimality,
and robustness against the modeling error
and the non-Gaussian noise stemming from the other societies' effects.
A Lyapunov stability argument was used to prove the boundedness of
the susceptible and infected populations to a small neighborhood around the origin. The convergence of the error solutions was also discussed under a proper selection of the robust gain. This
boundedness/convergence was achieved at minimal rates of the vaccination and antiviral treatment.
Simulation results illustrated that the proposed approach is able to provide accurate
state estimation, tracking performance, and robustness to the
modeling inaccuracies and the non-Gaussian noise associated with the nature of
the interactive human societies. This was achieved in an optimal control fashion. 

\subsection{Future works} \label{section.Conc3}
The control strategy developed in this study can be modified to be employed for a wide range of epidemiological diseases such as tuberculosis~\cite{Tube}, malaria~\cite{Mal}, Hepatitis C virus (HCV)~\cite{HCV-Sharifi}, HIV/AIDS~\cite{HIV-SMC}, and COVID-19~\cite{COVID-2}. In terms of future studies, the following items will be considered: 
\begin{enumerate}
\item In this paper, the system parameters $\Theta$ have to be guessed for use in the controller. However, to relieve the engineer of the need for such guess, future work is planned to design an adaptation mechanism to estimate these unknown parameters.
\item As illustrated in Figs.~\ref{fig.trackest},~\ref{fig.trackest50}, and~\ref{fig.trackestEKF}, the exposed population $z_2$ peaks at the beginning of the simulation. It implies that the number of people who are infected with influenza but not yet infectious initially increases and then vanishes as time goes on. Future work is planned to design a controller such that the exposed population is maintained below a number during the treatment period.
\end{enumerate}  

These items naturally encourage
us to extend the presented approach by estimating the system parameters and creating a safe control structure in which the exposed population is kept below a specified level.

%\addtolength{\textheight}{-0cm}
%%%%%%%%%%%%%%%%%%%%%%%%%%%%%%%%%%%%%%%%%%%%%%%%%%%%%%%%%%%%%%%%%%%%%%%%%%%%%%%%

\section*{References}
\bibliographystyle{model5-names}
\bibliography{Influenza}

\end{document}